# A review of cultural heritage inspection: Toward terahertz from mid-infrared region

Pengfei Zhu[1], Hai Zhang[2,1*], Stefano Sfarra[3], Dazhi Yang[4], Xavier Maldague[1]

[1]*Department of Electrical and Computer Engineering, Computer Vision and Systems Laboratory (CVSL), Laval University, Québec G1V 0A6, Québec city, Canada*
[2]*Centre for Composite Materials and Structures (CCMS), Harbin Institute of Technology, Harbin, 150001, China*
[3]*Department of Industrial and Information Engineering and Economics, University of L'Aquila, 67100, L'Aquila (AQ), Italy*
[4]*School of Electrical Engineering and Automation, Harbin Institute of Technology, Harbin 150001, China*

**Abstract**
This review examines non-invasive imaging methods spanning from the mid- and far-infrared up to the terahertz region (extending to about 1,000 μm) for the detection and analysis of Cultural Heritage artifacts. In the thermal infrared bands, where Planck's law governs radiation, the self-emission from materials can reveal intrinsic properties and internal degradation, whereas in the near-infrared, external illumination is essential to enhance surface detail and pigment differentiation. Similarly, far-infrared and terahertz techniques—employed in both transmission and reflection modes—provide complementary information by penetrating surface layers to expose underlying structures and hidden features. The full-imaging capability achieved by combining visible and infrared channels further enriches the diagnostic process by correlating traditional visual assessments with spectral data. In this review, in addition to introducing wide applications of these NII methods in cultural heritage, advanced signal processing techniques were summarized containing both hardware and software. Recent advances in deep learning have further revolutionized this field by enabling the classification, feature extraction, automated detection and super-resolution imaging of defects and degradation patterns. According to supervised and unsupervised learning, various neural networks can reliably identify subtle anomalies and material variations that may indicate prior restorations or early stages of deterioration. In conclusion, the synergy between advanced spectral imaging techniques, signal processing techniques, and deep neural networks promises a significant improvement in both the accuracy and efficiency of cultural heritage analysis, ultimately aiding in more informed conservation and restoration decisions.

## 1. Introduction

The preservation of Cultural Heritage has become an urgent global priority amid growing threats such as environmental pollution, urban development, natural disasters, and armed conflict [1]. As the value of safeguarding humanity's historical, cultural, and artistic legacy gains widespread recognition, non-invasive inspection (NII) techniques are attracting increasing attention for their ability to assess and monitor heritage materials without causing harm [2]. Compared to terms such as "non-destructive evaluation (NDE), non-destructive testing (NDT), and non-destructive inspection (NDI)", non-invasive inspection (NII) refers specifically to techniques that avoid both physical contact and any form of alteration, even minimal or reversible [3]. This stricter definition is particularly relevant in the context of cultural heritage, where the preservation of material integrity is paramount. Nevertheless, NII does not aim to replace the broader set of NDT techniques, many of which are still extensively and effectively used in heritage science. The choice between NII and other NDT approaches depends on various factors, including the nature of the object, conservation priorities, risk assessment, and available resources.

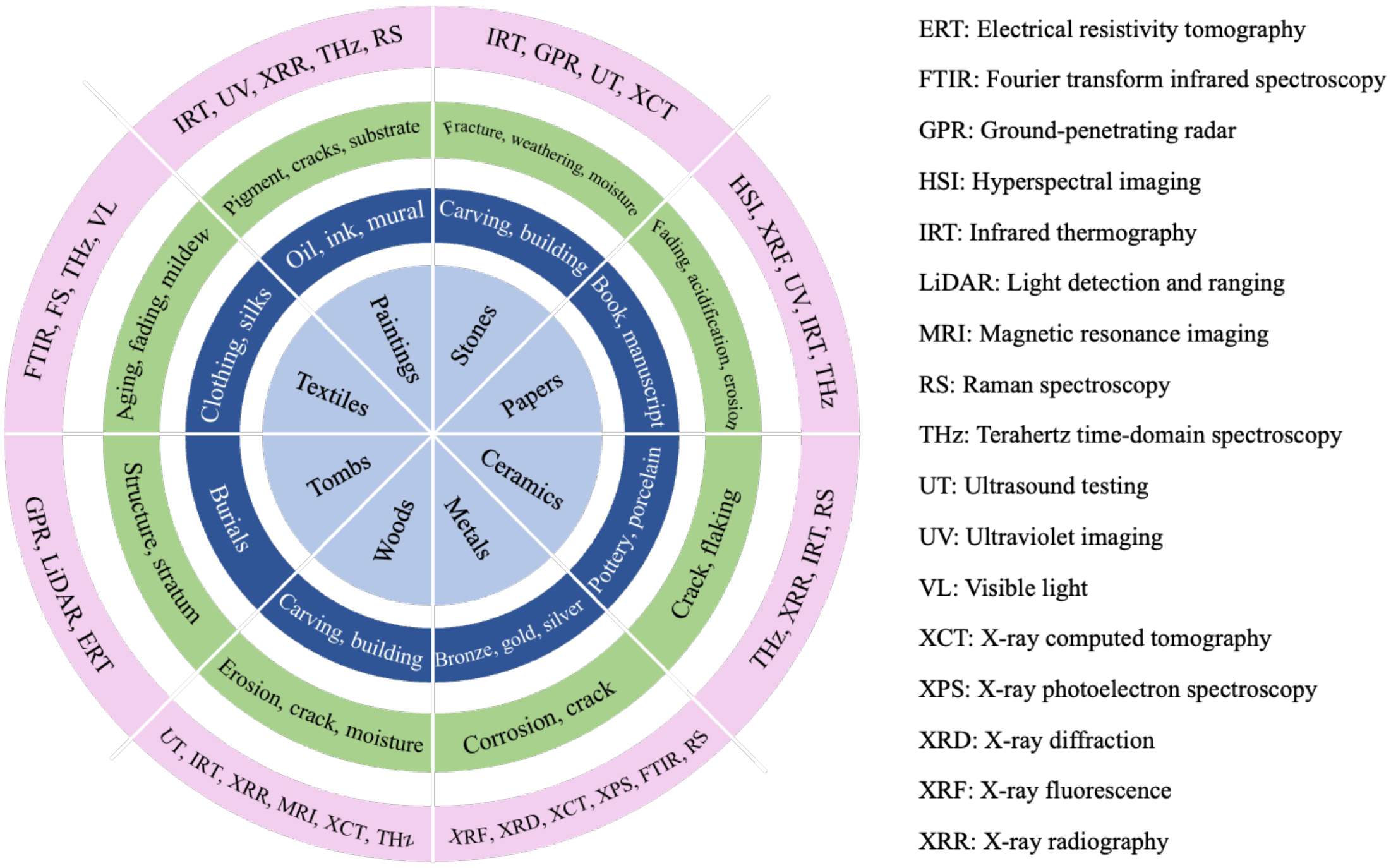

**Fig. 1.** The inspection of cultural heritage using different NDT techniques.

In Fig. 1, the authors summarized several NDT techniques and their corresponding application in cultural heritage which include organic and inorganic materials, divided into eight categories: (1) paintings, (2) stones, (3) metals, (4) ceramics, (5) papers, (6) textiles, (7) wooden artifacts, (8) buried sites. Diagnostic investigations play a key role in studying the material composition of artworks, their execution techniques, and their current state of preservation. As such, they provide crucial support for restoration work and form a fundamental basis for informed, sustainable, and reversible conservation strategies. The NDT techniques leverage principles from acoustics (ultrasound testing) [4], electromagnetics (X-ray radiography [5]. X-ray fluorescence [6], X-ray diffraction [7], X-ray computed tomography [8], X-ray photoelectron spectroscopy [9], terahertz time-domain spectroscopy [10], ground penetrating radar [11]), optics (visible light [12],

Fourier transform infrared spectroscopy [13], ultraviolet imaging [14], Raman spectroscopy [15], hyperspectral imaging [16], light detection and ranging [17]), electric (electrical resistivity tomography) [18], and magnetism (magnetic resonance imaging) [19], and thermal (infrared thermography) [20]. Compared with traditional NDT techniques, infrared thermography and THz-TDS attract increasing focus due to their contactless, non-ionized, and relatively low-cost properties. In this review, the authors focus on the application of NII techniques from mid-infrared to terahertz region, as shown in Fig. 2.

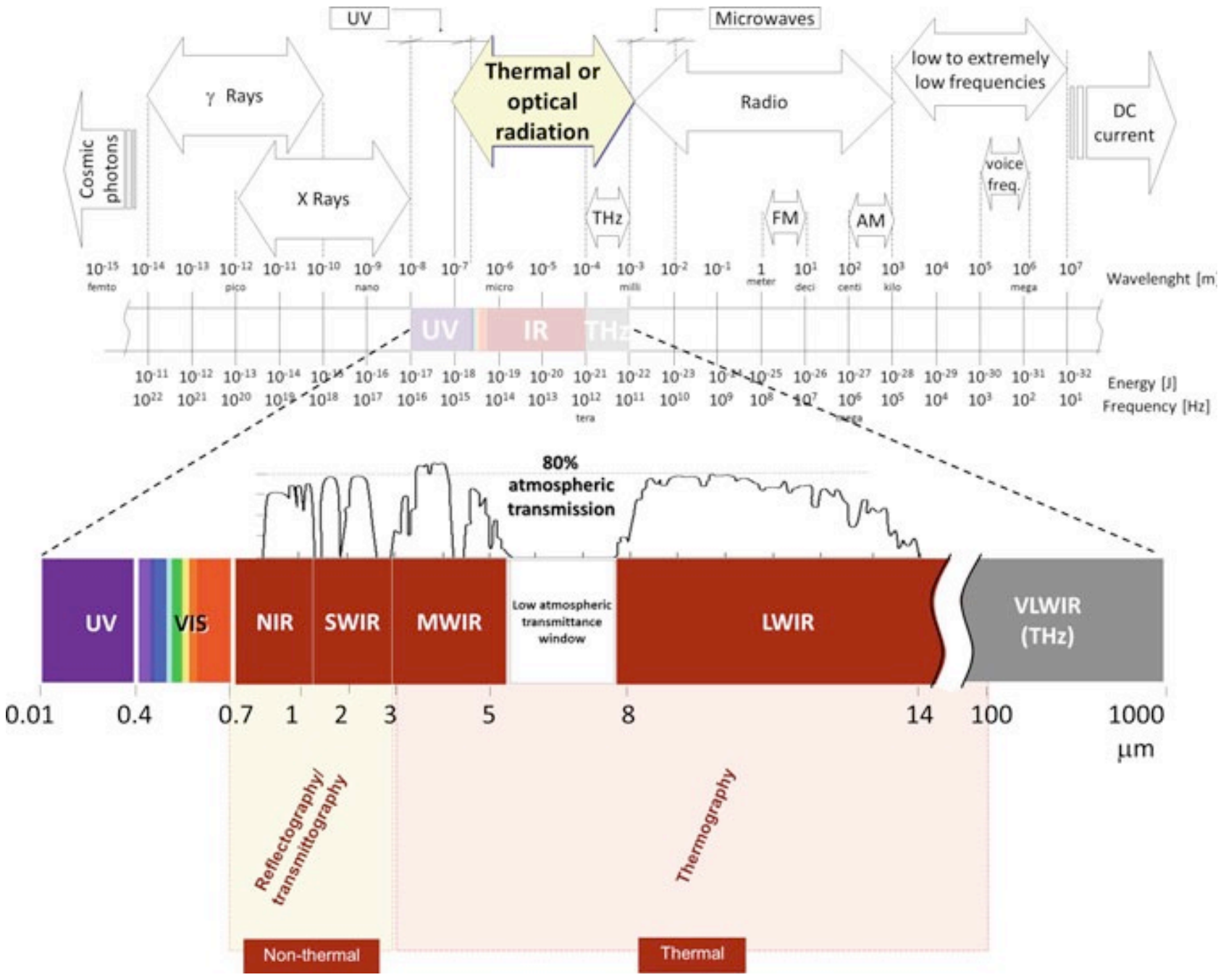

**Fig. 2.** Region of interest in the electromagnetic spectrum (from mid-infrared to terahertz).

According to their frequency (ν) or wavelength (λ), electromagnetic waves are categorized into several regions, including infrared (IR) and terahertz (THz). Infrared radiation is typically subdivided into [21]: Near-infrared (NIR, 0.75-1 μm), short-wave infrared (SWIR, 1-2.5 μm), mid-wave infrared (MWIR, 3-5 μm), long-wave infrared (LWIR, 7.5-14 μm). As frequency increases, photon energy rises according to $E=\hbar\nu$ (where $\hbar$ is the reduced Plank's constant), with photons exhibiting more pronounced quantum behavior. In contrast, THz radiation (0.1–10 THz, or 30 μm–3 mm) lies between the infrared and microwave domains and presents dual behavior, interacting both as electromagnetic waves and as quasi-particles. This allows THz waves to probe vibrational and electronic transitions, making them especially valuable for non-destructive analysis of stratified structures. It is worth noting that in active infrared thermography (AIRT) [22], the inspection relies on thermal excitation (heat input) rather than direct infrared light excitation, as the infrared radiation detected originates from the object's thermal emission. Terahertz time-domain spectroscopy (THz-TDS) operates on a similar principle [23], using short pulses of THz radiation to probe materials and obtain information on internal structures and low-frequency vibrational modes by analyzing the reflected or transmitted signals. Thermography was originally developed to visualize temperature distributions from a distance by detecting infrared radiation emitted by objects. However, the absorption of this radiation by atmospheric constituents can introduce measurement errors [24]. To minimize these effects, infrared cameras must operate within specific spectral windows that avoid major absorption bands, as illustrated in Fig. 3. These windows typically fall within the mid- and far-infrared regions. A summary of the key characteristics of these technologies is provided in Table 1.

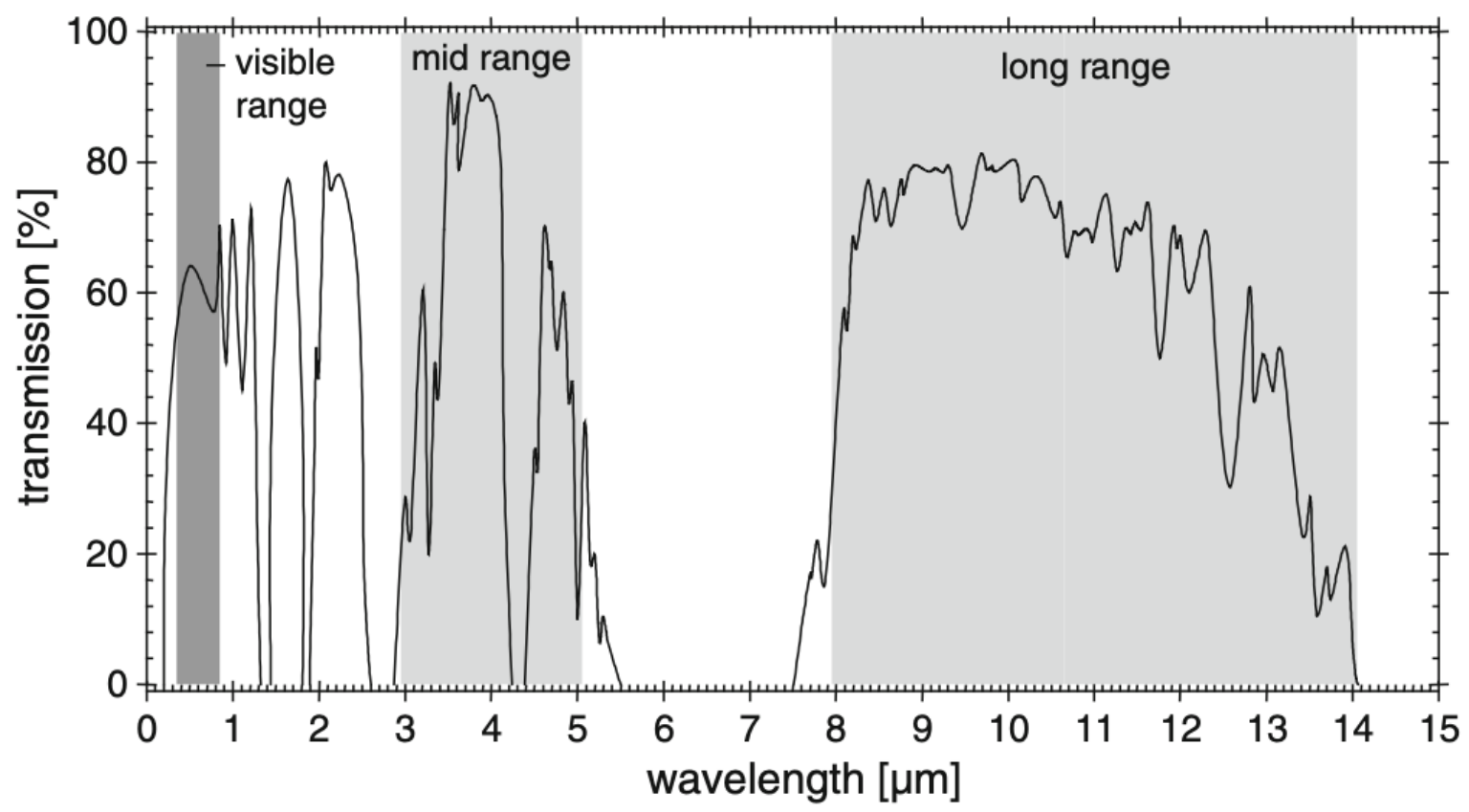

**Fig. 3.** Transmission of the atmosphere in the wavelength range between 0.2 and 14 μm. This measurement was done over a distance of 1250 m at 20 °C and 100% relative humidity [25].

**Table 1**

The features of infrared thermography and THz-TDS technology.

| System | SWIR camera | MIR camera | LWIR camera | THz-TDS |
|---|---|---|---|---|
| Wavelength | 1.4-3 μm | 3-5 μm | 7.5-14 μm | 30 μm-1 mm |
| Penetration | Shallow | Deeper | Deep | Deepest |
| Lateral resolution | High | Medium | Low | Lowest |
| Cost | High | Higher | Low | Highest |
| Imaging speed | Faster | Fast | Slow | Slowest |
| Detection goals | Subsurface defect | Subsurface defect | Subsurface defect | 3D tomography |

This review focuses on non-invasive inspection techniques for cultural heritage, spanning from the mid-infrared to the terahertz regions. The structure of the paper is as follows. Section 2 presents a review of key developments in infrared thermography, infrared spectroscopy, and terahertz imaging. Beyond describing each modality, their underlying principles and the main challenges encountered in practical applications are also discussed. Section 3 focuses on the implementation of these techniques across various categories of precious artifacts. Section 4 highlights recent advances in data processing and interpretation algorithms, while Section 5 explores the growing role of artificial intelligence in non-invasive diagnostics. Finally, Section 6 outlines several emerging techniques with strong potential for future applications. This critical and comprehensive review—spanning hardware, instrumentation, algorithmic frameworks, and practical implementations—aims to support the early-stage conservation and preservation of cultural assets.

## 2. Non-invasive inspection techniques: From mid-infrared to THz imaging

This section reviews several established infrared thermography and terahertz imaging techniques specifically developed for the inspection of cultural heritage. Both the underlying physical principles and experimental setups are introduced, along with the main challenges related to achieving high-resolution imaging and reliable quantitative analysis. Applications covering surface and subsurface detections—from the mid-infrared to the terahertz region—are discussed, with particular emphasis on the practical implementation of these methods in the investigation of authentic artifacts, rather than mockups.

### *2.1. Infrared Thermography Approaches: Passive and Active*

Infrared thermography approaches are generally classified as passive and active. The passive approach relies on natural thermal contrasts—typically due to environmental influences or solar radiation (solar loading)—to detect surface or subsurface anomalies without external stimulation. In the active approach, an external energy source is used to generate thermal contrast, enhancing defect visibility; common techniques include pulsed thermography, lock-in thermography, and pulsed compression thermography [26,27].

#### *2.1.1. Passive thermography*

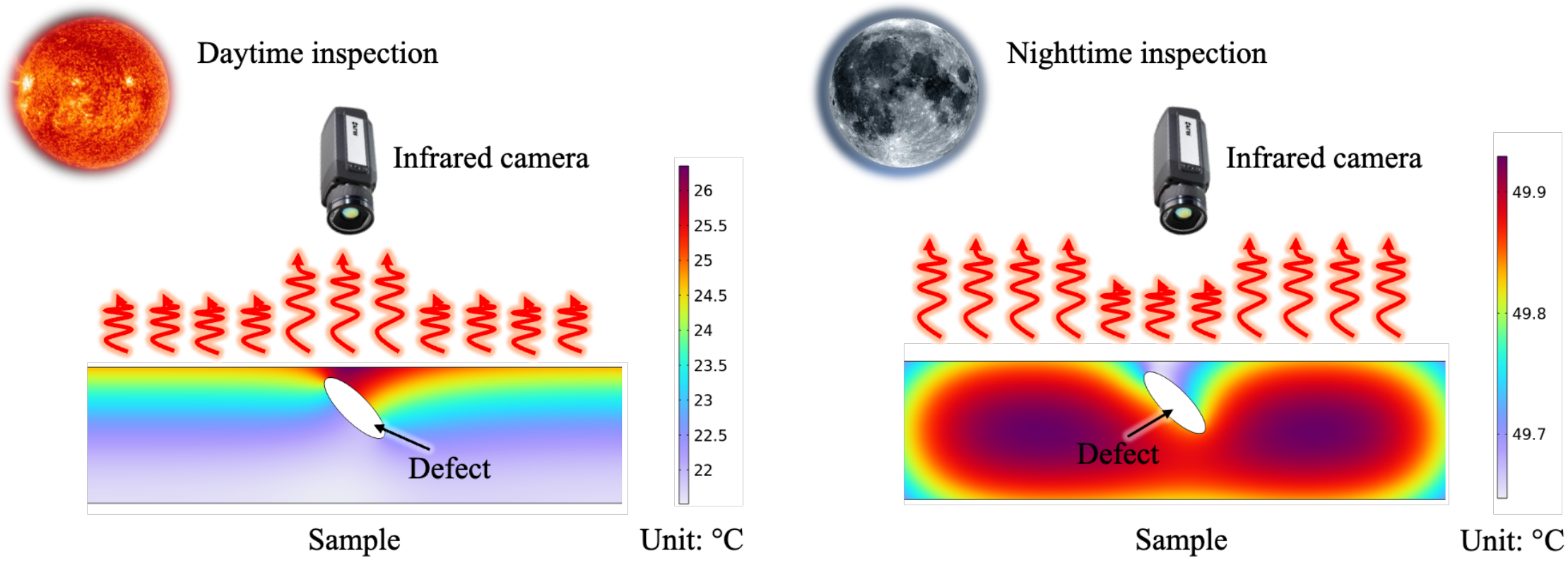

**Fig. 4.** Representation of principle of passive thermography inspections for samples with defects/voids.

Different from active infrared thermography, passive infrared thermography involves capturing thermal images of a structure without artificial external thermal stimulation, relying solely on the natural thermal emission from the sample's surface [28]. Fig. 4 shows the inspection of subsurface defects / voids using passive thermography. Passive infrared thermography involves capturing thermal images of an object without applying external heating, relying solely on its natural emission. Regions with lower thermal conductivity within the material alter the expected heat flow, generating contrast in thermal images. When the object is exposed to thermal loads—such as solar radiation during the day—heat penetration is hindered in these zones, leading to localized surface warming. At night, as the stored heat is released, the same areas exhibit delayed cooling, resulting in reduced infrared emission compared to the surrounding material.

The mathematic model for heat transfer is given by the energy conservation equation [29]:

$$\rho C_p \frac{\partial T}{\partial t} + \nabla \cdot (\mathbf{q} + \mathbf{q_r}) = 0 \tag{1}$$

where $\rho$ is the density, $C_p$ is the heat capacity at a constant pressure, $t$ is the time, $T$ is the temperature. The conductive heat flux is defined by Fourier's law as $\mathbf{q} = -k\nabla T$, with $k$ being the thermal conductivity. The operator $\nabla\cdot$ denotes the divergence. The term $\mathbf{q_r} = \varepsilon(G - \sigma T^4)$ approximates the net radiative heat flux, where ε is the surface emissivity, G is the irradiation, and σ is the Stefan-Boltzmann constant.
At the boundary, convective heat exchange with the environment is described by Newton's law of cooling [30]:

$$\mathbf{q_c} = h(T_{ext} - T) \tag{2}$$

where $h$ is the convective heat transfer coefficient and $T_{ext}$ is the ambient temperature. The detailed definition and modeling of the convective coefficient $h$ can be found in [31-33].

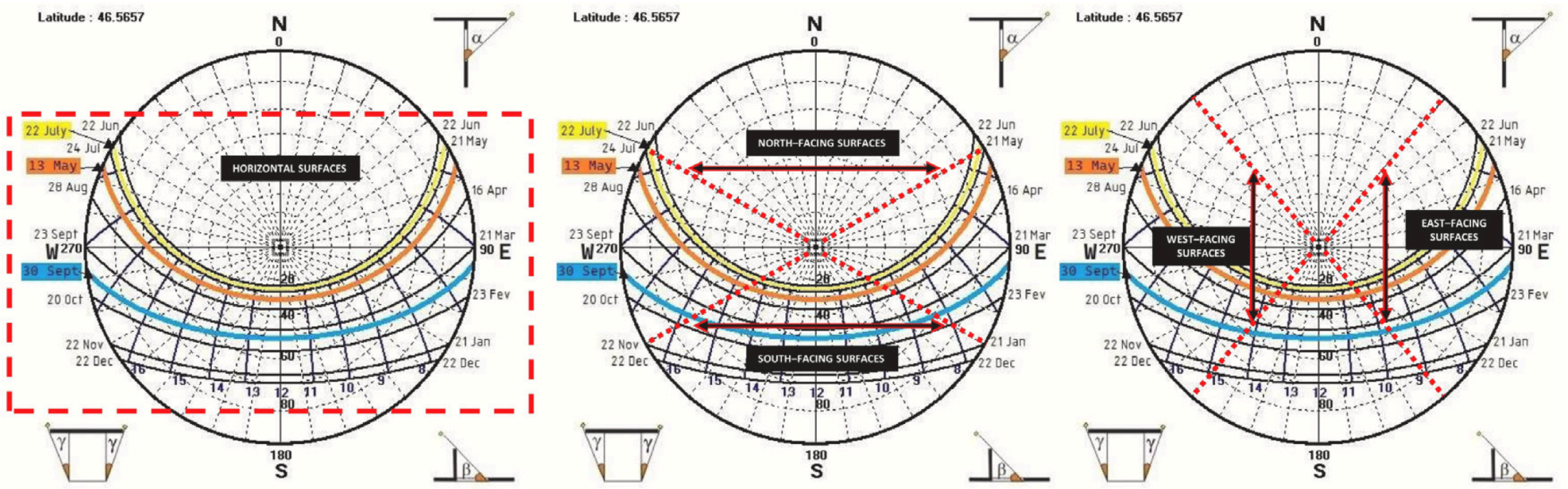

**Fig. 5.** Incidence of solar irradiation for the inspected surfaces at different days and times of data acquisition [38].

Thanks to controlled thermal stimulation, active infrared thermography generally offers higher sensitivity and specificity in defect detection and allows for the extraction of quantitative information through proper data analysis [34]. However, its application to large concrete structures poses significantly due to the energy demand and equipment required to uniformly heat or cool extensive areas uniformly. This limitation makes it less suitable for large-scale inspections, where passive thermography is often more feasible and cost-effective [35,36]; in such cases, quantitative evaluations are still when supported by thermal modeling [37,38]. Nevertheless, passive thermography also has limitations, primarily its strong dependence on environmental conditions during inspection [39-41]. For instance, its use is generally discouraged during the winter months [42,43]. Although year-round data acquisition is feasible, lower temperature differentials in winter reduce the effectiveness of the method. In particular, inspections should be avoided when ambient air temperatures fall below 0 °C, to prevent false readings caused by ice formation within surface anomalies [42]. Furthermore, understanding the solar trajectory at different latitudes is a critical step in passive thermography to select the right time and direction for observations, as shown in Fig. 5.

### *2.1.2. Pulsed thermography*

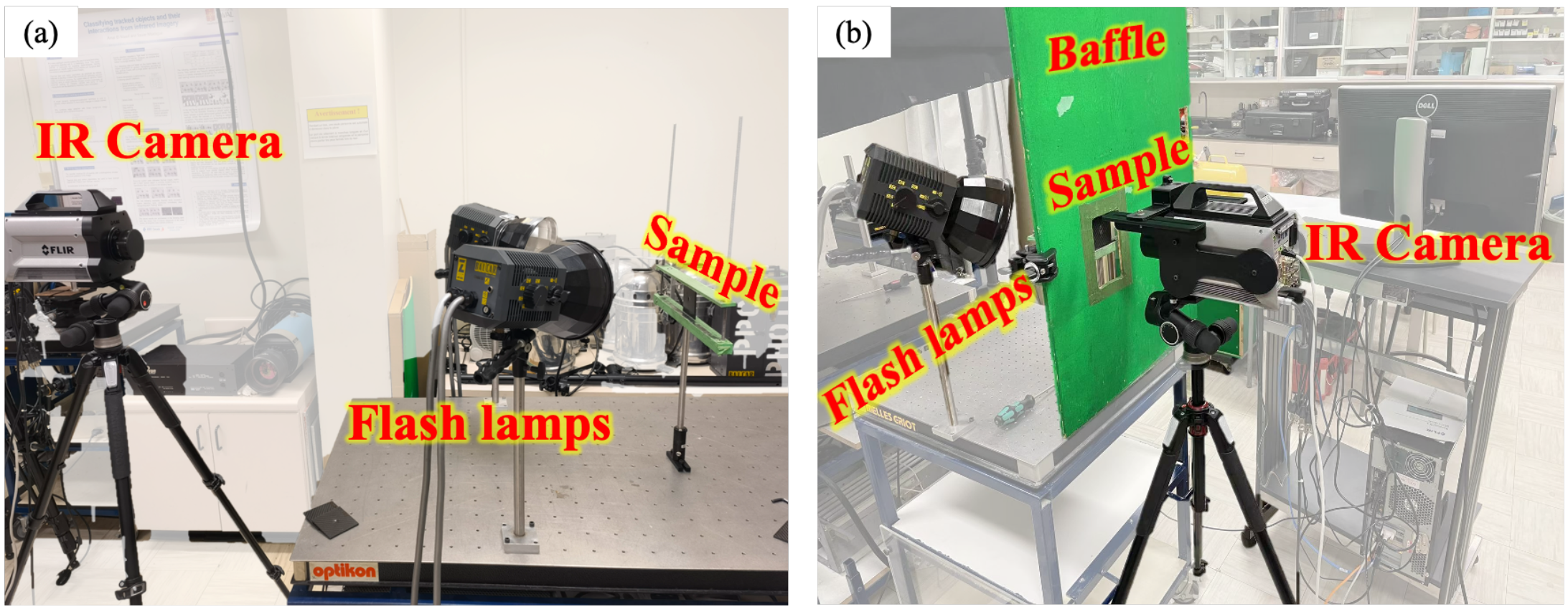

**Fig. 6.** Schematic images of pulsed thermography in (a) reflection mode and (b) transmission mode.

Pulsed thermography (PT) is largely used and perceived as the most sophisticated and versatile thermal NDT technique [44,45]. The process of pulsed thermography is illustrated in Fig. 6. The sample is thermally stimulated with a fast heat pulse (e.g., using xenon flashes) while the thermal evolution is recorded for as long as necessary so as to allow the sample to cool down [46]. Pulsed thermography can be applied in two different modes: the front face heating mode (reflection), where the heating and the detection are carried out on the same sample, and the rear face heating mode (transmission), where the heating and the detection are carried out on opposite sides of the sample [47,48]. It is worth noting that in both modes, the analysis of the time evolution of the temperature distribution at the sample surface allows researchers to quantitatively determine thermophysical properties (such as thermal diffusivity) of the material [49-52].

The one-dimensional heat conduction problem has been addressed by Carslaw and Jaegar [53]. The solution for a Dirac delta pulse plane applied at the surface $z$=0 of a semi-infinite solid (where z≫0) with a plane source of strength $Q/\rho C_p$ is given by:

$$T_{s-inf}(z,t) = \frac{Q}{\sqrt{\pi \rho C_p k t}} \exp\left(-\frac{z^2}{4\alpha t}\right) \tag{3}$$

where $z$ is the direction in which heat travels (depth), $\alpha$ is the thermal diffusivity, and $Q$ is the radiant energy. If the initial temperature distribution *T(z,0)* is known for a thermally insulated solid of uniform thickness $L$, the temperature distribution at time $t$ can be expressed as

$$T(z,t) = \frac{1}{L}\int_0^L T(z,0)dz + \frac{2}{L}\sum_{n=0}^{\infty} \exp\left(\frac{-n^2\pi^2\alpha t}{L^2}\right)\cos\left(\frac{n\pi z}{L}\right)\int_0^L \frac{T(z,0)\cos(n\pi z)}{L}dz \quad (4)$$

According to Eq. (4), Parker et al. [54] derived an analytical solution specific to pulsed thermography. When a pulse of radiant energy $Q$ is instantaneously and uniformly absorbed in the small depth $g$ at the front surface $z = 0$ of a thermally insulated solid of uniform thickness $L$, the temperature profile at that instant is given by

$$T(z,0) = Q/\rho C_p g \text{ for } 0 < z < g \quad (5)$$

and

$$T(z,0) = 0 \text{ for } g < z < L \quad (6)$$

Using this initial condition in Eq. (4), the transient temperature distribution can be written as

$$T(z,t) = \frac{Q}{\rho C_p L}\left[1 + 2\sum_{n=1}^{\infty}\cos\left(\frac{n\pi z}{L}\right)\frac{\sin(n\pi g/L)}{n\pi g/L}\exp\left(\frac{-n^2\pi^2}{L^2}\alpha t\right)\right] \quad (7)$$

Since $g$ is a very small for opaque materials, it's commonly assumed ~~follows~~ that $\sin\left(\frac{n\pi g}{L}\right) \approx \frac{n\pi g}{L}$. For PT in reflection mode ($z = 0$)

$$T(0,t) = \frac{Q}{\rho C_p L}\left[1 + 2\sum_{n=1}^{\infty}\exp\left(\frac{-n^2\pi^2}{L^2}\alpha t\right)\right] \quad (8)$$

and in transmission mode ($z = L$)

$$T(L,t) = \frac{Q}{\rho C_p L}\left[1 + 2\sum_{n=1}^{\infty}(-1)^n\exp\left(\frac{-n^2\pi^2}{L^2}\alpha t\right)\right] \quad (9)$$

It is important to note that the solution by Carslaw and Jaeger is derived for a semi-infinite domain, whereas Parker's formulation applies to a finite-thickness slab. It should be noted that PT has limited penetration depth (typically ~4 mm) compared to passive thermography, due to the short duration of the thermal excitation, which prevents deeper heat diffusion. The thermal diffusion length $\mu$ which depends on the angular frequency $\omega$ is given by $\mu = \sqrt{2\alpha/\omega}$. Furthermore, in the context of cultural heritage the complexity of materials—due to layered structures and variable thermal properties—can reduce the signal-to-noise ratio (SNR) and complicate the quantitative assessment of defect depth and size. Finally, thermal excitation may induce thermal fatigue or micro-damage, particularly in heat-sensitive materials such as pigment layers, varnishes, or ancient paper artifacts.

### *2.1.3. Other infrared thermography techniques*

In this section we discuss two active infrared thermography techniques used as well as other active thermography techniques and equally important in the field of cultural heritage. It includes lock-in thermography and pulse compression thermography [55,56], as shown in Fig. 7.

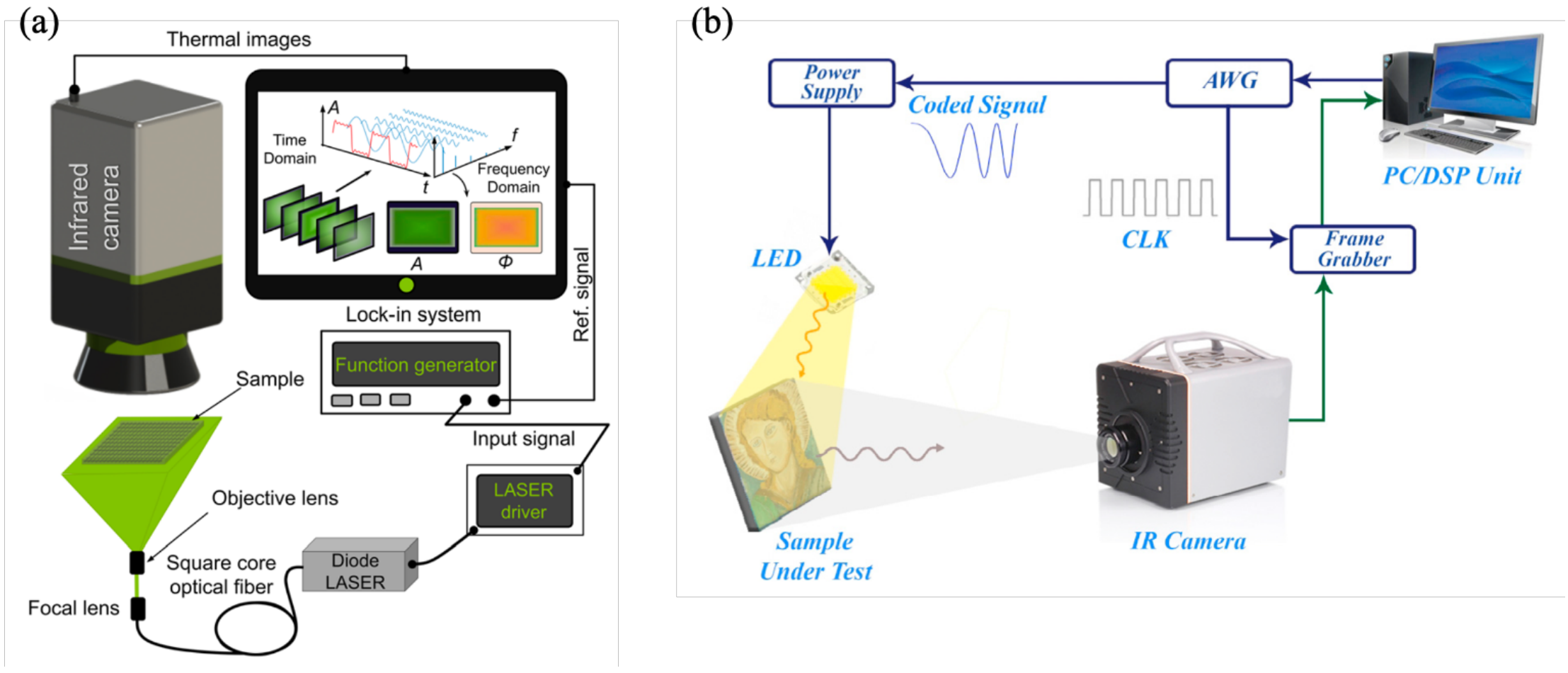

**Fig. 7.** Schematic images of (a) lock-in thermography [57] and (b) pulse compression thermography [58].

Lock-in thermography (LIT) is illustrated in Fig. 7(a). The system consists of an infrared camera, a function generator, and a lock-in processing unit. The function generator sends a periodic excitation signal at frequency $f$ to both the diode laser driver and the processing system. The laser beam, modulated accordingly, is expanded to ensure uniform heating and directed onto the surface of the sample. The resulting periodic temperature variations are captured by the infrared camera and analyzed by the lock-in system using the reference signal. The outcome is a set of amplitude ($A$) and phase ($\phi$) images extracted at the fundamental frequency via Fourier analysis [24]. The theoretical temperature distribution under harmonic excitation can be expressed as [57]:

$$T(z,t) = \frac{Q}{k\sqrt{\frac{2\pi^{3/2}f}{\alpha}}\sinh\left(\sqrt{\frac{2i\pi f}{\alpha}}d\right)}\cosh\left(\sqrt{\frac{2i\pi f}{\alpha}}(L-z)\right)e^{i\left(\omega t-\frac{3\pi}{4}\right)} \tag{10}$$

where $Q$ is the heat input, $\alpha$ is thermal diffusivity, $k$ is thermal conductivity, and $L$, $z$, and $d$ are geometric parameters of the sample.

Pulse compression thermography (PuCT), shown in Fig. 7(b), uses an arbitrary waveform generator (AWG) to emit a linearly modulated chirp signal and synchronize infrared image acquisition. The modulated signal drives a set of LEDs that apply coded thermal excitation to the sample. The theoretical temperature response is given by [59,60]:

$$T(z,t) = T_0 e^{2\pi i\left(f_0 t+\frac{Bt^2}{2\tau}\right)} e^{-z\sqrt{\frac{\pi}{\alpha}\left(f_0+\frac{Bt}{\tau}\right)}} e^{-iz\sqrt{\frac{\pi}{\alpha}\left(f_0+\frac{Bt}{\tau}\right)}} - \frac{2T_0}{\sqrt{\pi}} e^{2\pi i(f_0+Bt^2/2\tau)} \int_0^{z/2\sqrt{\alpha t}} e^{\frac{\pi i z^2}{2\alpha\mu^2}(f_0+Bt/\tau)} e^{-\mu^2} d\mu \tag{11}$$

where $T_0$ is the initial temperature, $f_0$ the starting frequency, $B/\tau$ the sweep rate, $B$ the bandwidth, $\tau$ the total duration, and $\mu = \sqrt{\alpha/\pi(f_0 + Bt/\tau)}$ is the thermal diffusivity.

Other techniques, such as square wave lock-in thermography [61], multi-frequency lock-in [62], vibrothermography [63], linear scanning [64], and eddy current thermography [65], are not discussed in this review due to the limited number of studies related to cultural heritage applications. In general, lock-in and pulse compression thermography offer greater penetration depth and higher signal-to-noise ratio compared to pulsed thermography [66]. However, they also require more complex setups, longer acquisition times, and more elaborate post-processing, which currently limit their applicability in real-time inspections. Moreover, both methods are sensitive to environmental and instrumental noise, which can affect the reliability of the results in practical scenarios.

### *2.2. Terahertz imaging*

For THz imaging, the following modalities are discussed: terahertz time-domain spectroscopy and continuous wave terahertz imaging.

#### *2.2.1. Terahertz time-domain spectroscopy*

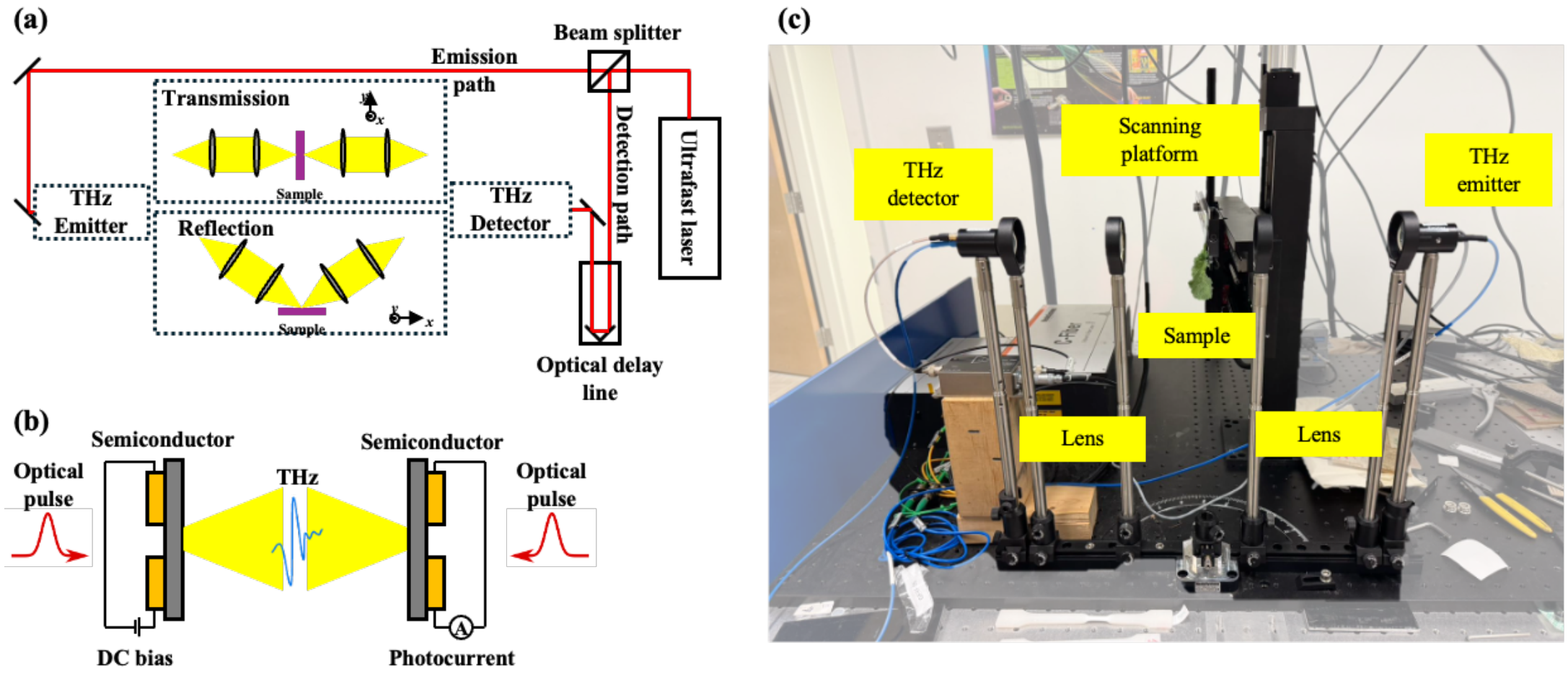

**Fig. 8.** THz-TDS systems: (a) schematic image of THz-TDS systems in transmission and reflection modes. (b) schematic image of photoswitches. (c) photo of THz-TDS system.

Historically, THz time-domain spectroscopy system (THz-TDS) has represented one of the earliest and most widely adopted techniques for the generation and detection of THz radiation [67]. In a typical THz-TDS system, broadband THz pulses are generated and detected using either photoconductive antennas or nonlinear optical processes. A photo-conductive antenna operates as a fast optically switch embedded within an antenna structure, as schematically illustrated in Fig. 8(a). This device generally consists of metal electrodes patterned onto a semiconductor substrate [68]. When a femtosecond optical pulse—whose photon energy approximately matches the bandgap of the semiconductor—is focused into the gap between the electrodes, it generates electron–hole pairs. These photo-generated carriers are then accelerated by a static electric field applied across the gap, as shown in Fig. 8(b). The transient current induced by the motion of these carriers results in the emission of a THz pulse. Following excitation, the carriers rapidly recombine or are trapped, allowing the system to return to equilibrium on a picosecond timescale. The photocurrent density $j_{em}(t)$ generated in the emitter can be expressed as the convolution of the optical excitation pulse with the impulse response of the photoswitch [69]:

$$j_{em}(t) = P_{opt}(t) * [n_{em}(t) q v_{em}(t)] \quad (14)$$

where $P_{opt}(t)$ is the incident optical power, and $q$, $n_{em}(t)$ and $v_{em}(t)$ are the charge, the density, and the velocity of photocarriers, respectively. A representative image of a complete THz-TDS set-up is shown in Fig. 8(c).

Compared with infrared thermography from mid-infrared to far-infrared regions, THz-TDS can provide higher SNR and deeper penetration depth (~ cm). However, this is a challenge that THz-TDS cannot penetrate samples with strong absorption in THz band including almost all conductors and partial liquids. Additionally, THz-TDS system relies on a point-by-point raster scan, which significantly limits the image acquisition time. The lack of economical and efficient THz sources and detectors remains an insurmountable problem in THz fields.

### *2.2.2. Continuous wave terahertz imaging*

The CW-THz system consists of a detector, a robot, and a CW-THz emitter. A compact sub-THz linear sensor array is based on the plasmonic detection method [70,71], which enables ultra-fast sensing of THz radiation at room temperature with response time of each plasmonic detector reaching 150 ps [72]. Each sensor is tuned to a single narrow band centered at 100 GHz. The sensor has a rectangular shape and 3 mm × 1.4 mm size. ~~In~~ A total of 256 sensors are arranged side-by-side lengthwise with 1.5-mm pitch to form a 1D array. Another distinctive component of the system is the IMPATT-100/80 wave generator-an extra compact, high-power sub-terahertz source that employs advanced IMPATT-diode technology. It is designed to supply a 100-GHz CW signal at $P > 80$ mW power level. The device is synchronized with camera, which makes possible lock-in like detection, and thus considerably reduces the total background noise of the camera sensors [73].

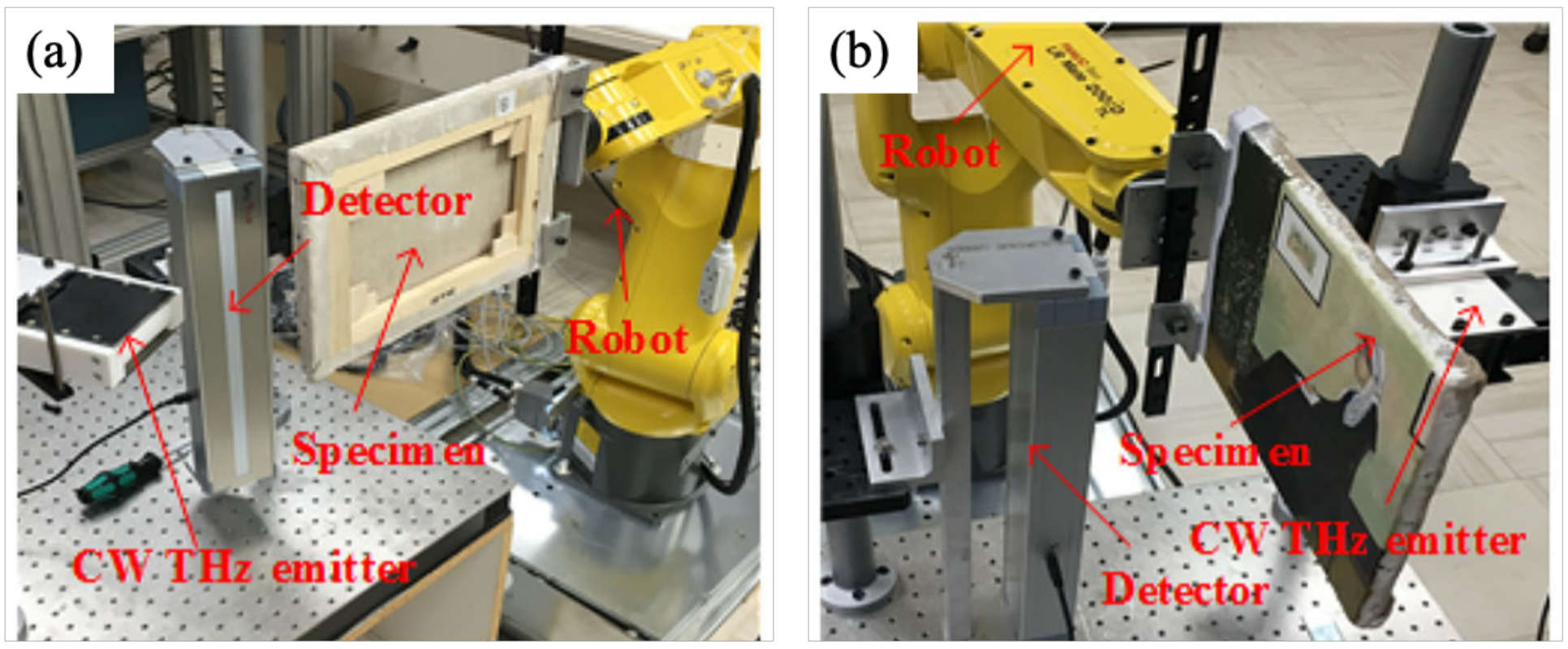

**Fig. 9.** CW-THz systems in (a) reflection mode and (b) transmission mode.

The aforementioned THz-TDS system not only relies on an expensive pulsed laser source, but it also requires an elaborated optical delay scheme based on free-space optics and optomechanics [73], cavity detuning [74], or a complex synchronization of two laser sources [75,76]. All three approaches pose high demands on assembly and adjustment or require complex electronic control. Optoelectronic continuous-wave (CW) THz system is an alternative due to its high spectral resolution [77]. The advantages of CW-THz systems are the unrestricted compatibility with optical fiber technology enabling fully fiber-coupled systems without any free space optics, moving parts or mechanical delay lines as well as the use of CW lasers instead of complex femtosecond pulsed lasers [78]. However, the output frequency of CW-THz systems is often limited below 1 THz. CW-THz is difficult to obtain the phase information as its lock-in like mode. Additionally, low time resolution, SNR, and system sensitivity also limit the further application of CW-THz systems. The comparison between THz-TDS and CW-THz systems are shown in Table 2.

**Table 2**

The comparison between THz-TDS and CW-THz systems.

| System | THz-TDS | CW-THz |
|---|---|---|
| Bandwidth | 0.1-10 THz | < 1 THz |
| Penetration depth | High | Low |
| Phase acquisition | Direct | Difficult |
| Cost | High | Low |
| Spatial resolution | Low | High |
| Imaging speed | Slow | Fast |
| Time resolution | Sub-ps | Low |
| Parameter Extraction | Strong | Weak |
| System complexity | High | Low |
| Cost | High | Low |
| SNR | High | Low |

### *2.3. Case studies of Infrared Thermography in Cultural Heritage*

Most studies on infrared thermography focus on the mid-infrared (MIR) region, as mid-wavelength and long-wavelength infrared cameras—commonly employed in practice—operate within this spectral range. Accordingly, the following section discusses the application of various thermographic techniques within the MIR domain.

Theodorakeas et al. [79] used passive thermography to inspect stone mosaics including the decorated floor discovered at the archaeological site of Delphi (Fig. 10(a)) and that can be dated back to the Early Byzantine Period, ruins of an ancient building (revealed during the excavation of Dörpfeld at the Ancient Agora of Athens and can be dated back to the 2nd century B.C.), and a decorative floor pattern realized through stone tesserae (found at the Sanctuary of Pan and can be dated between the 2nd century B.C. and 1st century A.D.). Stefano et al. [80] proposed a new excitation modality based on inside hot water stream for detecting a mosaic with artificial defects (Fig. 10(b)). As shown in Fig. 10(c), Spodek and Rosina [81] used solar loading to obtain a map of inner alterations including determining moisture diffusion in masonry of the Masonic Temple, built in 1926, identifying hidden defects and subsurface construction (Church of purification of St Mary, built between 1483 and 1500), detecting discontinuity between layers of plaster, both among themselves and with respect to the substrate, that usually preludes to the fall of the layers themselves of surface material (Old City Jail, built in 1802 until 1939). Ibarra-Castanedo et al. [82] conducted solar loading thermography (passive thermography) on north-west façade of Santa Maria della Croce di Roio church (L'Aquila, Italy). Researchers in National Technical University of Athens [83] applied passive thermography in large-scale conservation projects, such as the Pythian Apollo Temple in Acropolis of Rhodes, the Katholiko in Holy Sepulchre Temple, the National Archaeological Museum of Athens, and the Holy Aedicule of the Holy Sepulchre in Jerusalem (Fig. 14(d)). As shown in Fig. 14(e), Hatır et al. [84] employed passive thermography to monitor the capillary water rise, assess the cooling dynamics, and observe the drying behavior of saturated pyroclastic and travertine stones—both original blocks and restoration materials—used in the Kuruçeşme Han structure in Konya (Turkey).

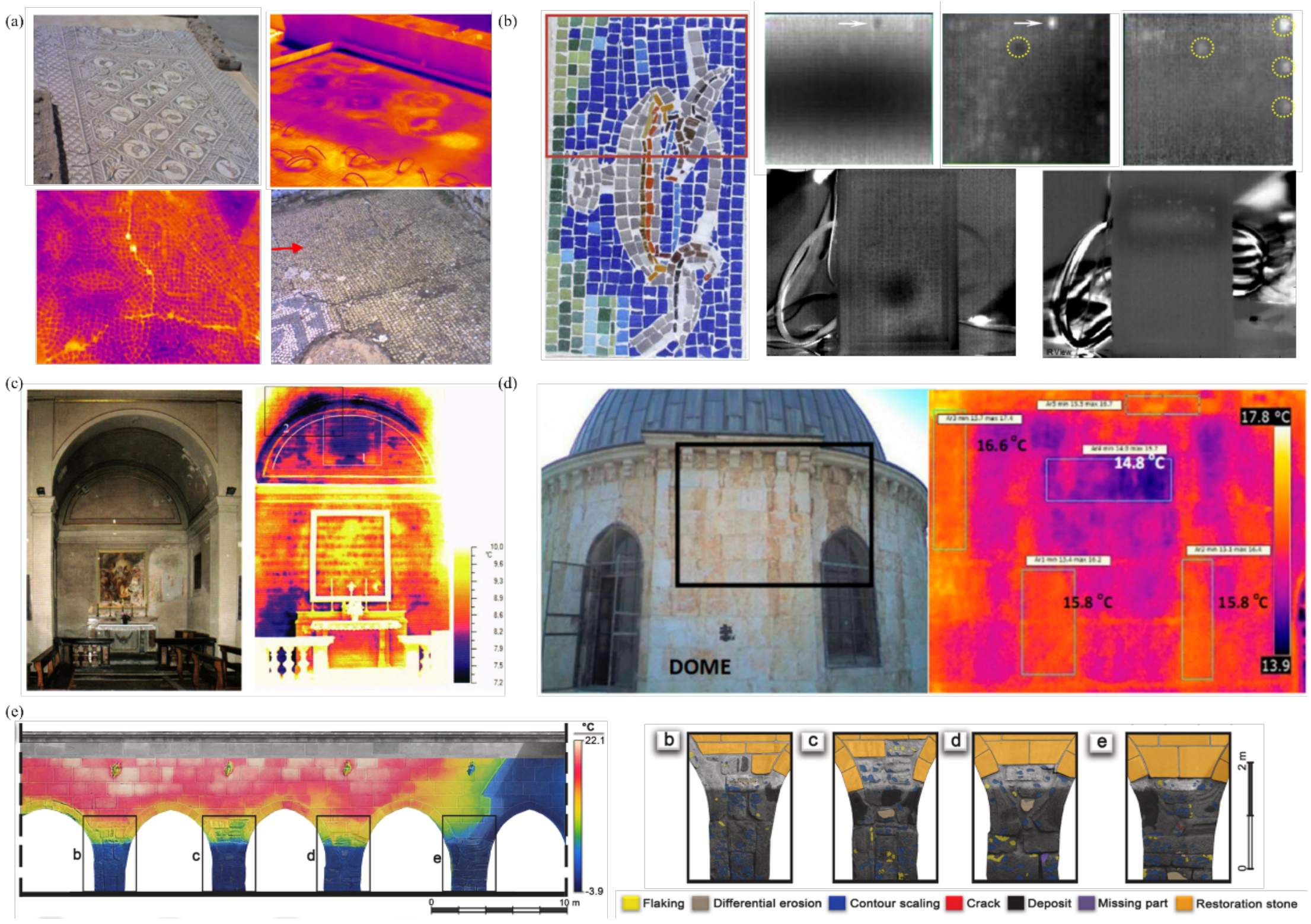

**Fig. 10.** Passive thermography for building inspection in MIR region: (a) Delphi floor mosaic's area and mosaic pavement at the Ancient Agora of Athens [79]. (b) Detection of mosaic with artificial defects using inside hot water stream excitation [80]. (c) Church of the Purification of St Mary, northern chapel. Caronno Pertusella, Italy [81]. (d) Dome of Katholiko in Holy Selphuchre Temple, Old City of Jerusalem [83]. (e) Kuruçeşme Han monuments, Turkey, built in 1207 (left – thermogram, right – deterioration map) [84].

Active thermography is particularly well suited for the analysis of small-scale and thin cultural heritage artifacts. Mercuri et al. [85] employed pulsed thermography to investigate both surface and subsurface features of the bronze statue Capitoline She Wolf, a symbolic artifact of the city of Rome (Fig. 11(a)). The same technique was subsequently applied to a 14th-century illuminated manuscript, the *Liber Regulare S. Spiritus de Saxia*, to reveal concealed elements beneath the painted layers (Fig. 11(c)) [86]. Ceccarelli et al. [87] integrated a mid-wave infrared (MWIR) camera with pulsed thermography and reflectography to detect hidden defects in the oil-on-canvas painting *La Primavera* by Mario Nuzzi and Filippo Lauri, (1658–1659) Chigi Palace, Ariccia (Italy) as shown in Fig. 11(b). A comparable diagnostic approach was adopted by Meo et al. [88] to examine a 16th-century panel painting by Marco Cardisco, *Adoration of the Magi*, Civic Museum of Castel Nuovo (Italy) (Fig. 11(d)).

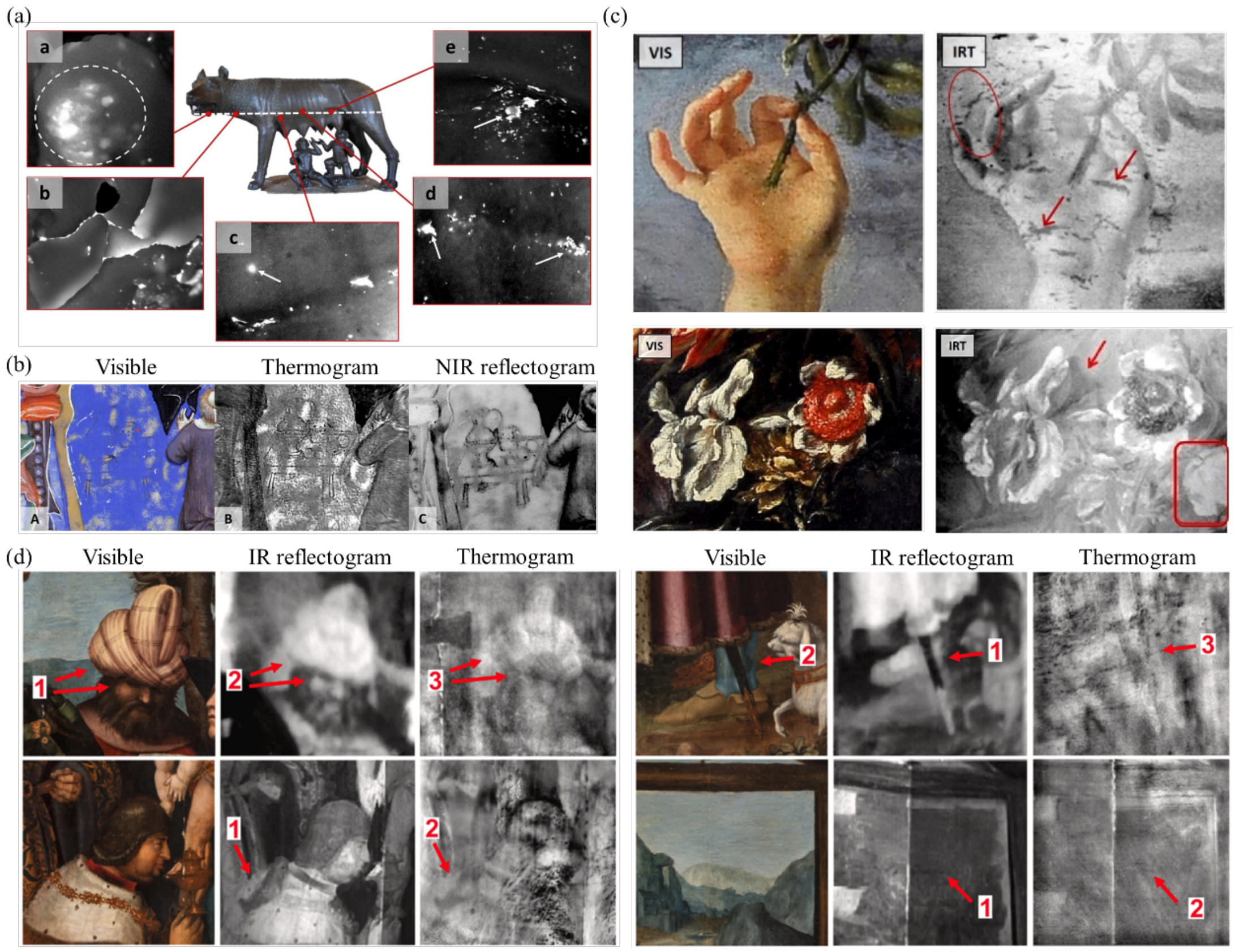

**Fig. 11.** Pulsed thermography in MIR region: (a) *Capitoline She Wolf* bronze statue, Capitoline Museums, Rome, Italy [85]. (b) Illuminated manuscript *Liber Regulae S. Spiritus de Saxia* of 14th century. Archivio di Stato di Roma (Italy) [86]. (c) *La Primave*ra, Chigi Palace, Ariccia (Italy) oil painting, made between 1658 and 1659 [87]. (d) *Adoration of the Magi*, panel painting of 16th century Civic Museum of Castel Nuovo (Italy) [88].

Hillen et al. [89] compared lock-in thermography and pulsed thermography in mid-wave and long-wave infrared for the inspection of three paintings, which include a 19th century Russian icon that depicts Saint Nicholas of Myra, a reduced-size copy of a self-portrait of Rembrandt of 1633 (now in the collection of the Louvre Museum, Paris), and a 20th century icon painting depicting St. George slaying the dragon (Fig. 12(a)). Maierhofer et al. [90] used lock-in thermography to detect a 2 cm thickness crack in a historical masonry of church of St. John the Baptist (Slovenia) which is composed of several materials: brick, stone, mortar, plaster, wood, metal (Fig. 12(b)). Calicchia et al.

[91] applied lock-in thermography to characterize the deterioration of painted surfaces within the delicate environment of the Greek chapel in the Priscilla catacombs (Fig. 12(c)).

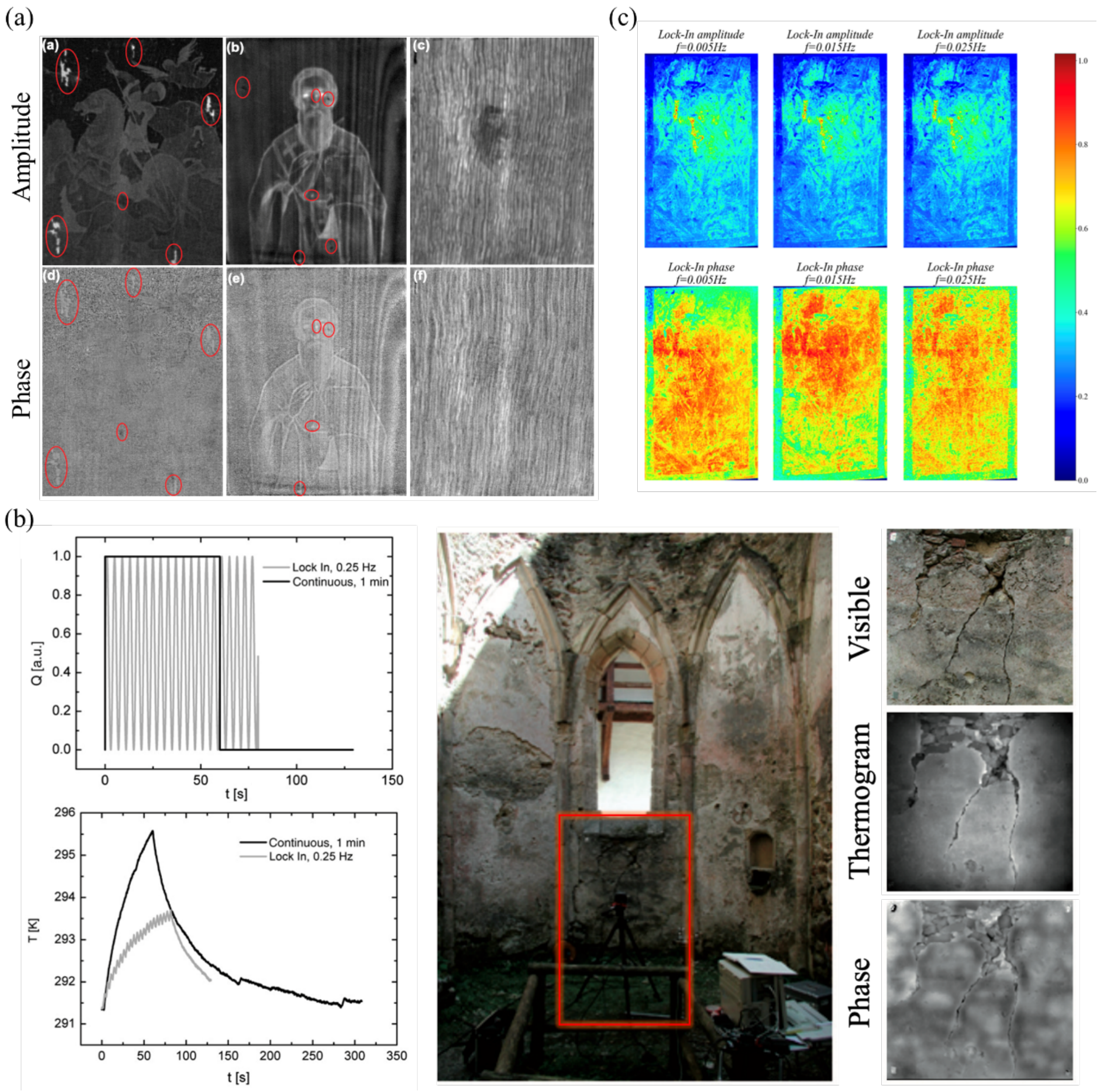

**Fig. 12.** Inspection using lock-in thermography: (a) Three paintings of *St. George*, *St. Nicholas*, and *Rembrandt*, Louvre Museum, Paris [89]. (b) Church of St. John the Baptist at the Carthusian monastery at Žiče, Slovenia (upper left – excitation signal, lower left – measured temperature signal) [90]. (c) The Priscilla catacombs (Rome, Italy) excavated between 2nd and 5th centuries [91].

Ricci et al. [92] combined use of the hypercolorimetric multispectral imaging (HMI) and pulse-compression thermography on a 15th century wall painting *Madonna and the Child enthroned* attributed to the Italian artist *Antonio del Massaro* (Fig. 13(a)). Vettraino et al. [93] used pulse-compression thermography to detect cracks and discontinuities in the ground layers of a 16th century wall painting within Palazzo Gallo in Bagnaia (Viterbo) and evaluate the depth of such anomalies, giving valuable support in the consolidation step (Fig. 13(b)). Laureti et al. [94] presented the contextual use of PuCT and HMI for the diagnostic study of *The Crucifixion* of Viterbo (Italy) attributed to Michelangelo's workshop, 16th century painting panel in the Museum of Colle del Duomo in Viterbo, Italy and a late 15th century painting panel preserved at Museo Carrara in Bergamo, Italy (Fig. 13(c)).

In addition to the aforementioned thermographic techniques, the eddy current pulsed thermography (ECPT) [95] was employed to detect surface cracks and defects in metallic artifacts (Fig. 13(d)). It is noted that the ECPT is only suitable for conductors, and its heat diffusion is from inside to outside.

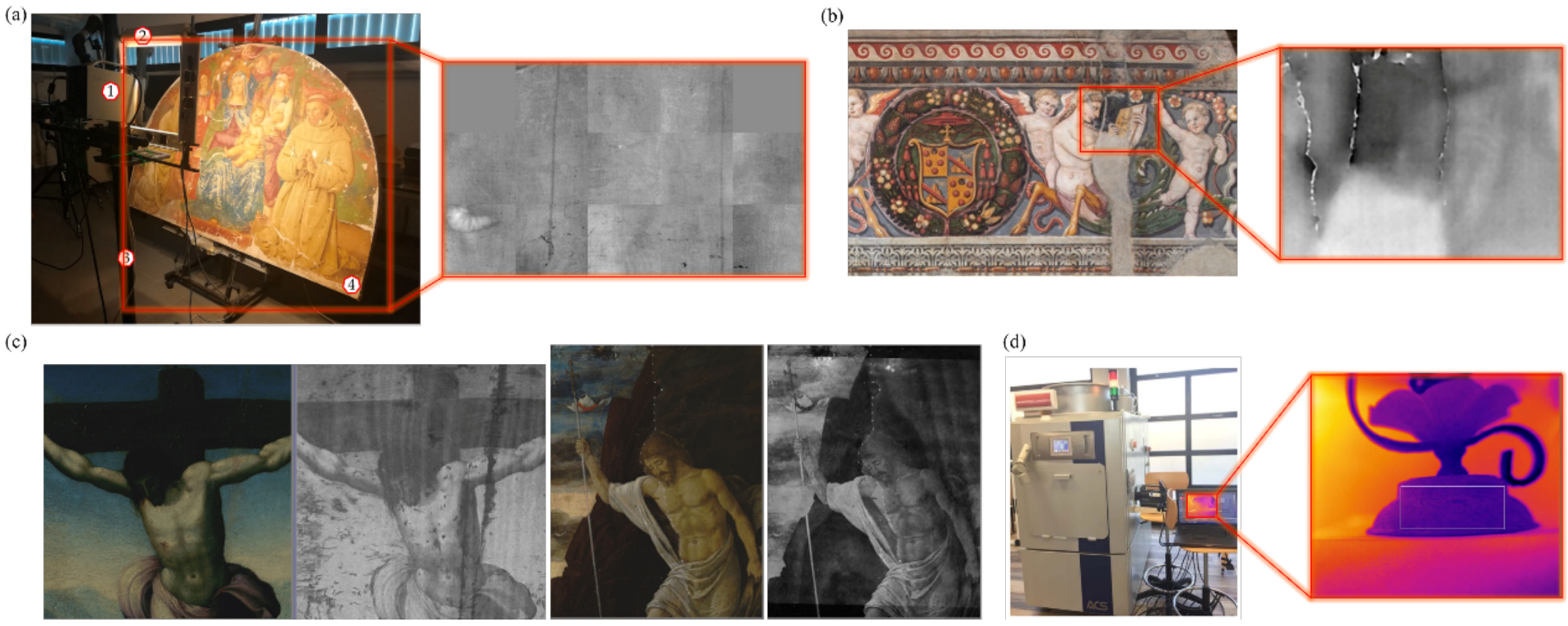

**Fig. 13.** Inspection using pulse-compression thermography (a-c) and eddy current pulsed thermography (d): (a) Wall painting dated back to 1490 from *Antonio del Massa*ro. Civic Museum of Viterbo, Italy [92]. (b) Wall painting in Palazzo Gallo (Viterbo, Italy) built within 1500-1520 AD [93]. (c) Painting panel of 16th century, Viterbo (Italy), and 15th century, Bergamo (Italy). HMI UVF on the left and PuCT on the right [94]. (d) Ancient oil lamp dated back to 1800s [95].

It should be noted that the thermal excitation used in infrared thermography may induce potential damage, including thermal stresses from overheating and the degradation of heat- and light-sensitive materials such as pigments and surface coatings. To mitigate these risks, Shrestha et al. [96] recommended conducting numerical simulations prior to experimental testing.

## *2.4. Case studies of THz-TDS in Cultural Heritage*

Terahertz imaging techniques represent a non-destructive testing technique as they are contactless and employ non-ionized radiation. According to the formula $E = \hbar v$, for THz photons in the frequency range of 0.1-10 THz, carry energies of approximately 0.4-40 meV. In contrast X-ray photons in the 0.1-10 EHz range possess energies in the order of 0.1-100 keV. The significantly lower energy of THz photons prevents ionization of molecular structures, making them inherently safe for sensitive materials.

The first application of A terahertz time-domain spectroscopy (THz-TDS) system was used to wall painting mock-up was reported study wall painting replicas for the first time by Jackson et al. in 2008 [97] (Fig. 14(a)). Subsequently, Giovannacci et al. [98] applied terahertz time-domain imaging (THz-TDI) technique to immovable cultural heritage, which is a specifically an ancient wall painting (Fig. 14(b)). Fukunaga et al. [99] employed used THz reflection imaging to detect characterize crack depth in a historic mural painting within a Lamaism temple (Inner Mongolia, China) (Fig. 14(c)). These studies demonstrate that THz spectroscopy and imaging can provide complementary information to conventional multi-spectral approaches in cultural heritage diagnostic. Inuzuka et al. [100] applied THz-TDI to the Takamatsuzuka mural paintings, (Japan) (Fig. 14(e)) revealing not only the internal stratigraphy of the plaster layers but also providing insights into the constituent materials of the artwork. Similar investigations focused on mural paintings can be found in [101,102] (Fig. 14(e) and (f)).

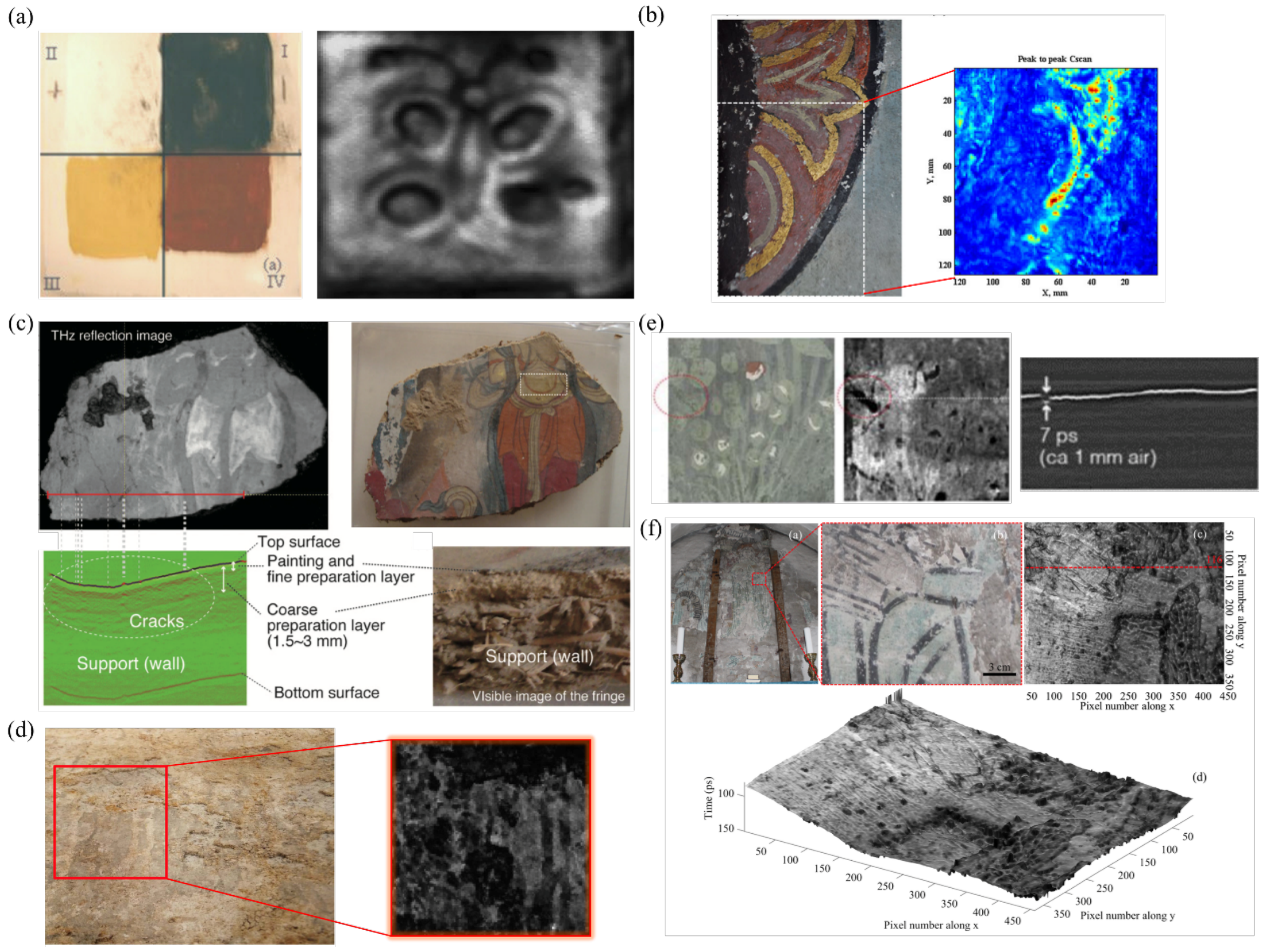

**Fig. 14.** Inspection of wall paintings using THz-TDS: (a) The butterfly is extracted from between plaster layers [97]. (b) THz topography of the wall painting [98]. (c) Wall painting from Dazhao Monastery, (China) [99]. (d) Takamatsuzuka Kofun Murals paintings, (Japan) [100]. (e) Apsidal wall painting in Nebbelunde Church, (Denmark) [101]. (f) Wall painting Annunciation fresco at the San Marco Museum, Florence (Italy) [102].

For-heritage inspection and documentation, Vazquez et al. [103] proposed a super-resolution THz-TDS systems to examine a medieval parchment fragment, dated back to 14th century A.D. (Fig. 15(a)). Krügener et al. [104] evaluated the internal damage state of two works of art, a medallion from the Castle of Celle (North of Germany) and a windowsill from the St. Peter of Trier Cathedral (Germany) using THz-TDS (Fig. 15(b)). Dandolo et al. [105] developed an integrated data processing framework combining THz-TDI and spectral-domain optical coherence tomography (SD-OCT), enabling the generation of co-registered cross-sectional images within a unified spatial grid. This approach was applied to the inspection of a painting mock-up realized with multiple layers showed in Fig. 15(c). Additionally, Fukunaga et al. [106] designed a compact THz imaging system for the analysis of art materials.

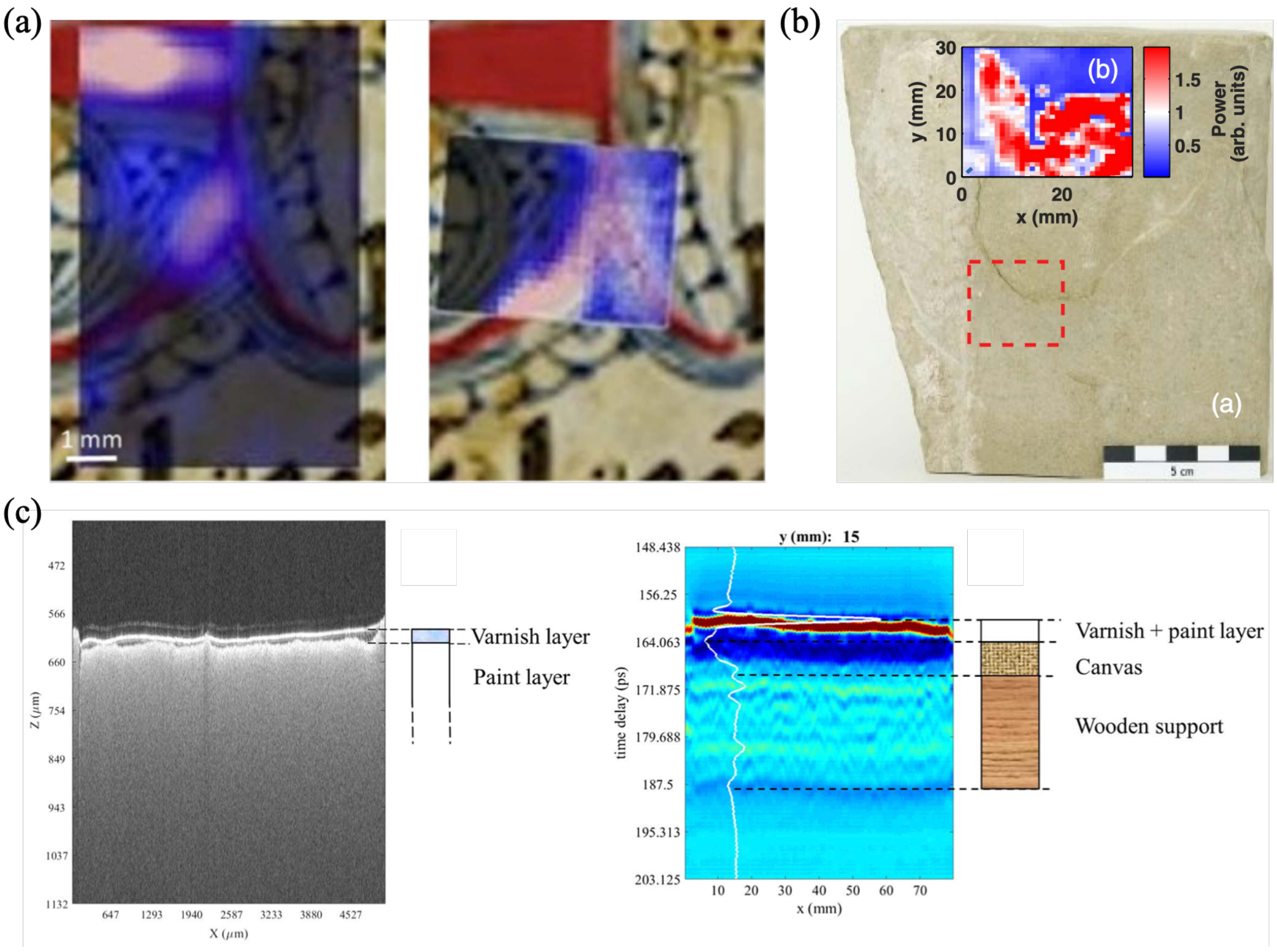

**Fig. 15.** Inspection of documentation: (a) Medieval parchment fragment dated back to 14th century AD [103]. (b) Fragment of the window sill removed from the St. Peter of Trier Cathedral (Germany) [104]. (c) Painting on canvas mockup with a thin wooden panel support [105].

The studies mentioned above illustrate the application of terahertz time-domain spectroscopy systems to cultural heritage inspection. In the following section, applications based on continuous-wave terahertz systems will be presented.

Zhang et al. [107] employed a continuous-wave (CW) THz imaging system operating at 0.1 THz to examine painting on canvas both in reflection and in transmission modes (Fig. 16(a)). Due to low operating frequency, the resulting images exhibited a low signal-to-noise ratio. Bauer et al. [108] presented frequency-modulated continuous-wave (FMCW) terahertz measurements of Leonardo da Vinci's wall painting (Italy): *The Last Supper* to investigate the sub-surface structure of the masterpiece (Fig. 16(b)). Ciano et al. [109] proposed a compact sub-THz confocal microscope operating at 0.3 THz and applied it to the analysis of a Russian wooden icon of an unknown craftsman dated back between the 17th and the 19th century (Fig. 16(c)).

In addition to 2D terahertz imaging, terahertz-computed tomography (THz-CT) has emerged as a valuable technique for revealing internal material structures. For the first time, Abraham et al. [110] applied the THz, X-ray, and neutron computed tomography to an art-related object (Fig. 16(d)), i.e., an Eighteenth Dynasty Egyptian sealed pottery preserved at the Museum of Aquitaine (Bordeaux, France). Wang et al. [111] applied

THz-CT technique to inspect samples oil painting mockups or folded paper and scrolls (Fig. 16(e)).

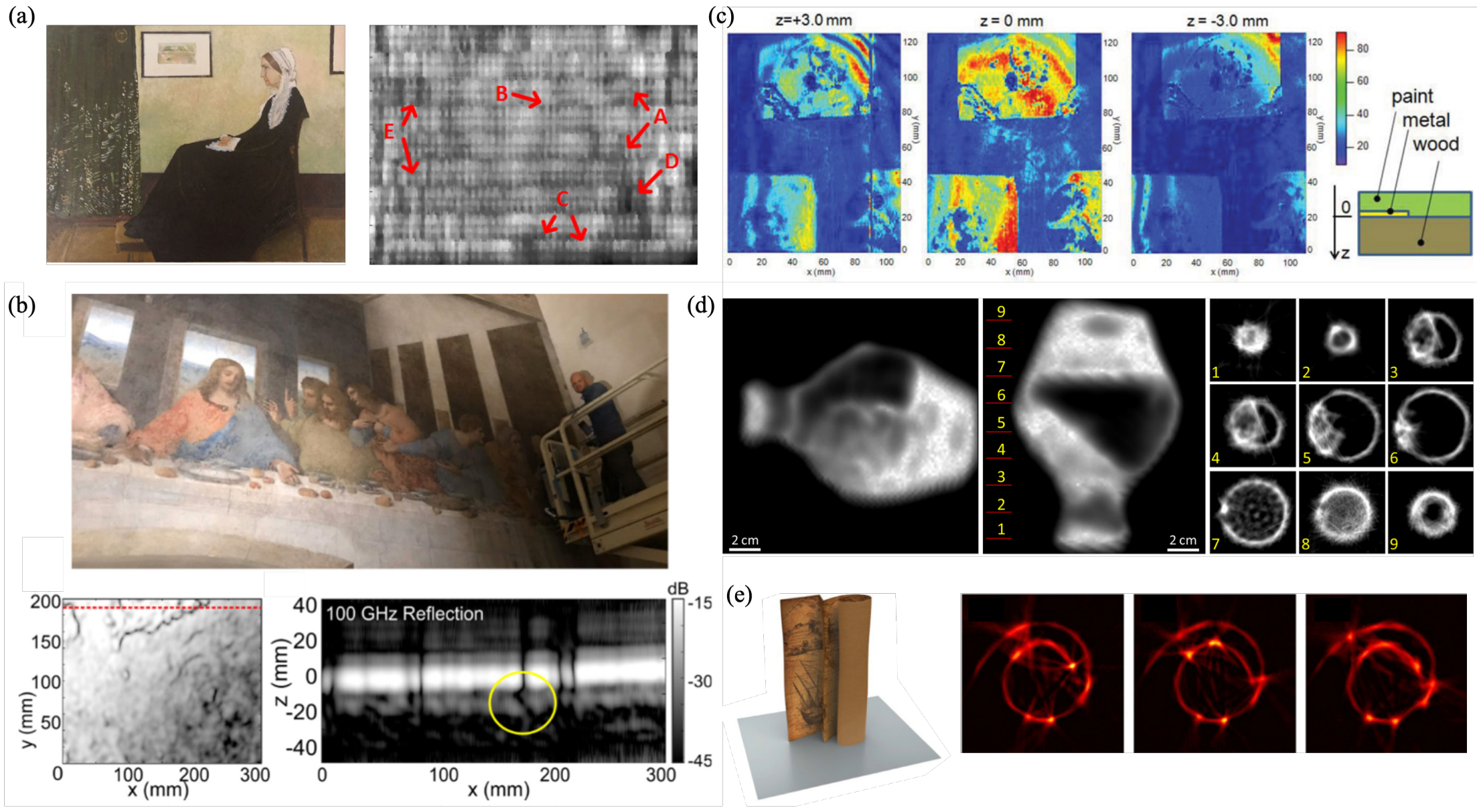

**Fig. 16.** Inspection of using CW-THz and THz-CT techniques: (a) Oil on canvas *Portrait of the Painter's Mother* Musée d'Orsay, Paris [107]. (b) *The Last supper* of Leonardo da Vinci at the convent of Santa Maria delle Grazie in Milano, Italy [108]. (c) Russian wooden icon dated back between the 17th and the 19th century [109]. (d) Egyptian pottery vase at Museum of Aquitaine (France) [110]. (e) Mockup of a folded document [111].

## 3. Signal processing algorithms for IRT and THz techniques

### *3.1. Infrared thermography*

If reviewing the development history of signal processing algorithms in IRT, it is necessary to mention four classic but effective methods, i.e., differentiated absolute contrast (DAC) [112], temperature signal reconstruction (TSR) [113], pulse phase thermography (PPT) [114], and principal component thermography (PCT) [115]. The summary of these methods is shown in Table 3.

Differentiated absolute contrast (DAC) method is derived from the absolute temperature contrast method [116], which calculates the temperature difference $T_{AC}$ between all pixels $T$ and a specific pixel $T_s$ at sound area

$$T_{AC}(t) = T(t) - T_s(t) \tag{15}$$

There are two assumptions in absolute temperature contrast method. First, the location of sound area is pre-known. Second, there is no uneven heating. However, such knowledge about the location of sound area is not always available while the thermal stimulation is barely uniform in case of PT experiments. Pilla et al. [112] proposed a modified version of absolute temperature contrast method known as differentiated absolute contrast (DAC). In DAC, it is considered entire the sample surface acts as non-defective for a time, $t^*$, which is an instance between the time of heat pulse launch and the time of appearance of the first defect on the thermogram. Considering the thermal evolution for semi-infinite body, then

$$\Delta T_s(t) = \Delta T(t^*)\frac{\sqrt{t^*}}{\sqrt{t}} \tag{16}$$

**Table 3**

Classic signal processing algorithms in IRT.

| Algorithms | DAC | TSR | PPT | PCT |
|---|---|---|---|---|
| Technical means | Time-domain denoising | Time-domain denoising | Frequency-domain analysis | Dimensionality reduction |
| Scope | Pulse thermography | Pulse thermography | All but most suitable for PT and LIT | All |
| Advantages | Easy to implement | Reducing data storage | Depth-related | (Almost) Best processing effect and highly adapted for all IRT techniques |
| Disadvantages | Low robustness | Low effect | Requiring high captured accuracy of infrared camera | High computational cost, low interpretability |

where $\Delta T_s$ and $\Delta T$ are the temperature increase from $t = 0$ at sound area and arbitrary area, respectively. Considering initial temperature of $T(0)$ and replacing Eq. (16) into Eq. (15)

$$T_{DAC}(t) = \big(T(t) - T(0)\big) - \big(T_s(t) - T_s(0)\big)$$

$$= \Delta T(t) - T(t^*)\frac{\sqrt{t^*}}{\sqrt{t}} \tag{17}$$

The accuracy of DAC method will be reduced as the time elapses and as plate thickness increases. Benitez et al. [117] proposed a corrected form of DAC based on Laplace transform to address this problem. However, this corrected DAC needs pre-known parameter of sample thickness and physical properties.

Temperature signal reconstruction (TSR) method is proposed by Shepard et al. [113]. It is based on the high order polynomial curve fit over the logarithmic temperature data

$$\ln[T(t)] = \textstyle\sum_{n=0}^{N} a_n[\ln(t)]^n \tag{18}$$

For a defect free sample, Eq. (18) results a straight line with slope of -1/2. The curve will deviate if there is a defect existing. Besides reducing the noise, TSR method can effectively reduce the data storage by several fitting parameters. Additionally, Zhu et al. [118] proposed an improved TSR method for long pulse thermography.

Pulse phase thermography (PPT) introduced by Maldague and Marinetti [114] is the most used signal processing algorithm in IRT fields. It combines the advantages of PT and LIT. PPT technique uses the unscrambling feature of discrete Fourier transform (DFT) in order to decompose the thermal signal to its amplitude and phase modules. A pulse thermal response can be transformed from time domain to frequency domain

$$F_n = \sum_{k=0}^{N-1} T(k) \exp\left(-\frac{i2\pi kn}{N}\right) = Re_n + iIm_n \quad (19)$$

where *Re* and *Im* denote respectively the real and imaginary parts of transformation, *n* denotes the frequency increment, and *N* is the frame number. The calculation of amplitude and phase information can be given as

$$Am_n = \sqrt{Re_n^2 + Im_n^2} \quad (20)$$

$$\varphi_n = \tan^{-1}(\frac{Im_n}{Re_n}) \quad (21)$$

Besides the effect of signal processing, the physical meaning of PPT is quite significant. The penetration depth of thermal wave is related to the modulated frequency ($\mu = \sqrt{\alpha/(\pi f)}$). In the case of lock-in thermography, there is only one modulated frequency in the calculated spectrum. However, PPT technique can provide a wide spectrum including different frequency components, which corresponds to different depth information.

Principal component thermography (PCT) technique was proposed by Rajic [115]. Principal component analysis works by first calculating the covariance matrix of the original variables. The covariance matrix describes the relationship between the variables and shows how much they vary jointly. If two variables are highly correlated, their covariance is large in magnitude, whereas if they are not correlated, their covariance is zero. The covariance matrix is then decomposed into a set of orthogonal eigenvectors and their corresponding eigenvalues. The eigenvectors represent the directions in the original variable space that explain the largest amount of variation in the data. The eigenvalues represent the amount of variance explained by each eigenvector. The eigenvectors are sorted in descending order of their corresponding eigenvalues, with the first eigenvector explaining the most important variability and the last eigenvector explaining the least [119]. Considering a three-dimensional array $T(i, j, k)$, where $i$=1,2,…,$N_x$, $j$=1,2,…,$N_y$, $k$=1,2,…,$N_t$, it can be reduced to a two-dimensional array by a raster-like operation on the pixel values in each image frame, producing a matrix A with dimension $(N_xN_y)\times N_t$. Then a singular value decomposition (SVD) is performed on matrix *A*

$$S = (A - A_{mean})(A - A_{mean})^T = U\Gamma U^T \quad (22)$$

where $S$ is the covariance matrix, U corresponds to the matrix containing the eigenvectors of $S = (A - A_{mean})(A - A_{mean})^T$, $A_{mean}$ denotes the mean temporal profile, $G$ is a diagonal matrix containing the singular values of $S^T = (A - A_{mean})^T(A - A_{mean})$ eigenvalues. Comparing with DAC, PPT, TSR algorithms, PCT technique has better signal processing effect, which has been validated by some literatures [120-122]. In addition, PCT technique is not only suitable for PT experiments, but also for all kinds of IRT experiments. Subsequently, Yousefi et al. [123] proposed the application of candid covariance-free incremental principal component analysis in thermography, which demonstrated computational efficiency and an incremental, covariance free version of the original PCT method. Furthermore, other approaches include non-linear principal component analysis, known as sparse principal component analysis [124], and low-rank sparse principal component thermography [125]. Lopez et al. [126] applied a statistical correlation method, partial least squares regression (PLSR), to experimental PT data from a carbon fiber-reinforced composite with simulated defects, showing that PLSR has a similar effect to PCT. Principal component analysis and independent component analysis (ICA) are component analysis approaches, which means that both try to identify the most meaningful basis to project a multivariate dataset from a mathematical space to another. The ICA attempts to decompose the dataset as a set of independent signals. It is often used for blind signal separation. Morabito et al. [127,128] were among the first to use INCA with eddy-current testing (ECT) initially to analyze their features and later to detect and segment defects. Shin et al. [129] used ICA to identify defects from an ECT distorted signal. Yang et al. [130] investigated the ICA application for defect detection in pulsed eddy-current (PEC), which was later improved by Cheng et al. [131].

Infrared thermography is typically considered as a 2D imaging technique. However, researchers tried to explore the relationship between depth and time evolution. In 1990, Vavilov proposed a dynamic thermal tomography (DTT) technique [132]. However, low signal-to-noise ratio limits the development of DTT technique. In 2014, Kaiplavil and Mandelis [133] introduced truncated-correlation photothermal coherence tomography (TC-PCT), founded on optical excitation thermography utilizing a pulse chirp signal. However, the TC-PCT method significantly relies on high frame-rate (~ms) sampling of the infrared camera, which leads to the requirement for computer support with increased data storage capabilities. Recently, Zhu et al. [134] proposed a frequency multiplexed photothermal correlation tomography (FM-PCT) technique. The FM-PCT improves depth resolution and defect localization precision despite the diffusive nature of the probing (photo)thermal waves, using an infrared camera with lower frame rate than TC-PCT (as low as 50 frame/s) and extends the excitation to linear heating sources. As a result, it offers faster sampling speeds and calculation time, and can be used with more economical mid-IR cameras making it more widely available.

### *3.2. Terahertz time-domain spectroscopy*

The most signal processing research was focused on 1D THz signals because the reflected THz wave can carry the depth information of the inner boundaries. This property makes reflected THz pulses ideal for imaging multi-layer structures and bulks with inside voids and defects [135,136]. The reflected time-domain signal comprises multiple time-shifted and amplitude-attenuated pulses, which are formed by the temporal waveform of the incident pulse convoluted with the impulse responses of the interfaces [135], therefore the impulse-response function of the system can be estimated by deconvoluting the time signal by the temporal waveform of the incident pulse.

However, the echoes reflected by the interfaces overlap if the layer or defect is too thin, and the echoes may submerge by noise in poor detection.

Numerical deconvolution method was first investigated to map the distribution of layers for single-layer and multi-layer paint films [137]. A method based on multiple regression analysis was later proposed to improve the resolution at the expense of signal-to-noise ratio (SNR) [138]. Researchers attempted to realize higher resolution of thickness measurement by compressing the pulse width [139]. However, the physical limit of the pump source is constrained. Subsequently, researchers focused on filtering out the noise spikes in the higher frequency region of the impulse-response function by thresholding [140]. Further study was employed to signify the impulse response by combining Wiener deconvolution and wavelet shrinkage [141]. This approach was named hybrid frequency-wavelet domain deconvolution (FWDD). However, the impulse responses retrieved by FWDD are broadened by the cutoff of the low-SNR region, which limits its resolvability for overlapped echoes [142].

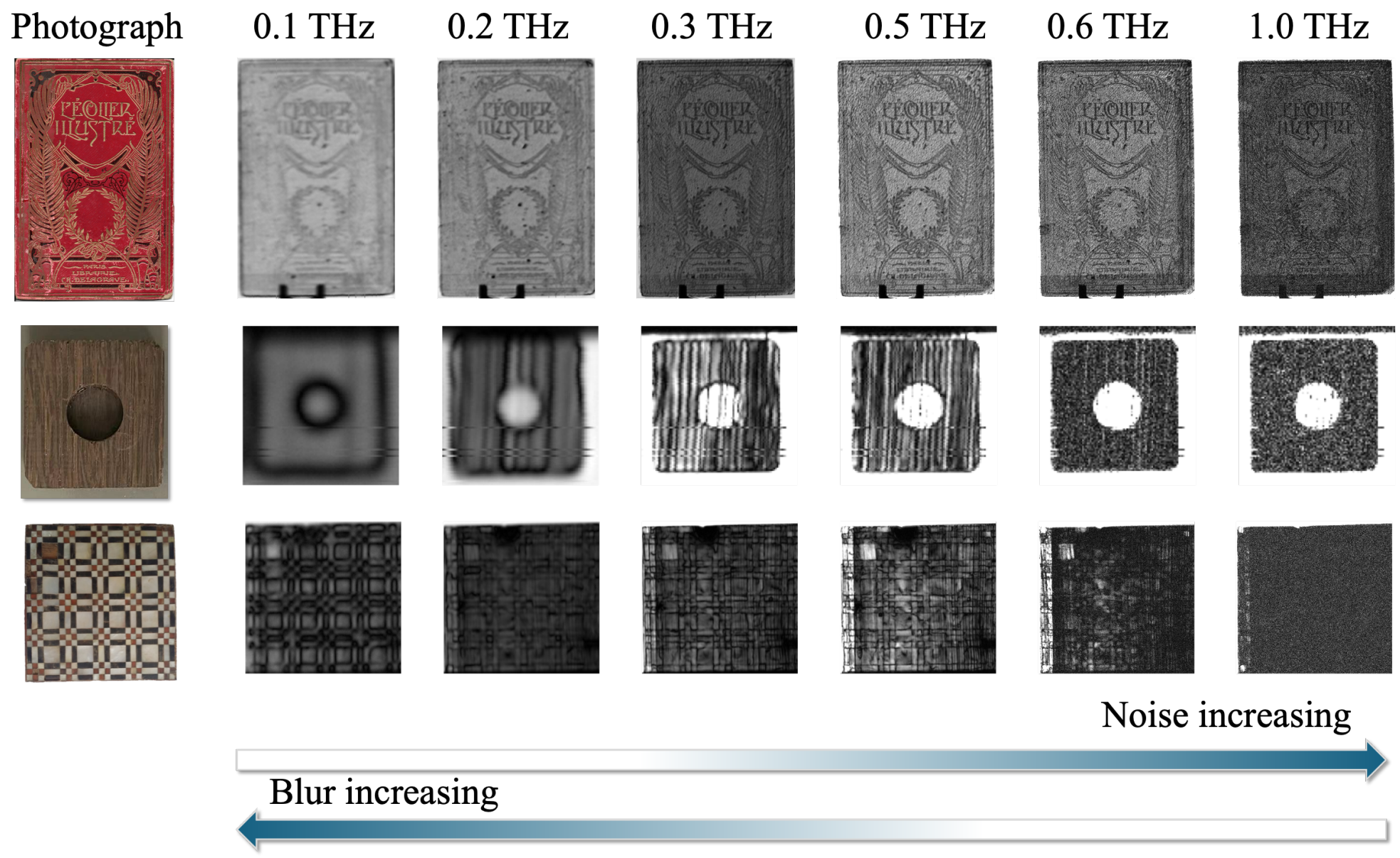

**Fig. 17.** Typical degradation effects in the representation of the amplitude images. The images are corrupted by noise that increase with the frequency, while at lower frequency value, the blur effects are more dominant. The samples contain a book cover in [119], a pyrolysis wood in [145], and a marquetry in [187].

The design of the THz system, including the focusing optics and beam-forming process, introduces frequency-dependent degradation effects that distort the amplitude images. The primary factors contributing to the degradation and subsequent loss of spatial information (Fig. 17) in THz images include: 1) the shape of THz beam; 2) system-dependent noise; 3) losses resulting from THz wave propagation through the sample, such as reflection and/or refraction; 4) the scattering and absorption properties of the sample. While the latter two factors are inherently sample-dependent and difficult to model accurately, the effects of the THz beam shape and system noise can be estimated using simple approximations.

There are few literatures regarding 2D THz image processing. Xu et al. [143] applied projection onto a convex sets approach, iterative backprojection approach, Lucy-Richardson iteration, and 2D wavelet decomposition reconstruction to THz imaging for obtaining a high-resolution THz image with decreased noise. Ljubenović et al. [144] proposed a joint deblurring and denoising approach for restoring high-resolution hyper-spectral images in THz-TDS. Zhu et al. [145] proposed a normalized time-domain integration method to extract important spectral information.

In terahertz (THz) imaging, enhancing image resolution and reducing signal acquisition time have been major areas of interest compared to other image processing objectives. Terahertz time-domain spectroscopy (THz-TDS) systems offer ultrafast temporal resolution at the sub-picosecond scale, enabling the acquisition of both amplitude and phase information over a broad THz frequency range by capturing the temporal response of an object. However, due to the single-pixel nature of current THz-TDS systems, image data are typically obtained through raster scanning of the object or the imaging system, leading to slow imaging speeds and bulky, complex setups.

To eliminate the need for raster scanning, some researchers have explored electro-optic processes in nonlinear crystals (NLC) to convert the THz amplitude and phase information of an object into the near-infrared (NIR) range, which can then be detected using optical cameras [146-149]. However, the nonlinear nature of the wavelength conversion process requires high-power, expensive laser sources, and results in a low signal-to-noise ratio (SNR). Another approach to avoiding raster scanning involves using time-varying spatial modulators (TVSM) to encode THz radiation patterns interacting with the object and reconstructing the image based on spatial mode information [150-154]. This method eliminates the mechanical scanning platform, making the THz imaging system faster and more stable. However, these imaging systems require multiple time intervals to encode THz radiation patterns with different spatial distributions, which limits their speed, especially when acquiring image data over a broad THz bandwidth. Additionally, the need for time-varying spatial modulators increases the system's size, complexity, and cost. Terahertz focal-plane arrays (THz-FPAs) offer a promising alternative by providing simultaneous spatial, ultrafast temporal, spectral, amplitude, and phase information of an object, potentially revolutionizing THz imaging technology. However, existing THz detector arrays based on field-effect transistors [155] and microbolometers [156] lack temporal and frequency resolution and do not capture phase information. To push the spatial resolution of ultrafast THz spectroscopy beyond the diffraction limit and into the critical length scales of materials science, various imaging concepts have been developed. Techniques such as aperture-coupled detectors (ACD) [157], tapered photoconductive probes (TPP) [158], laser terahertz emission microscopy (LTEM) [159], and near-field electro-optic sampling have achieved spatial resolutions on the order of a few micrometers [160]. However, all THz microscopy techniques that have achieved spatial resolutions below 100 nm rely on coupling THz radiation to sharp metallic tips. The THz field is enhanced and confined within the length scale of the tip apex diameter, forming the basis for scattering-type scanning near-field optical microscopy (s-SNOM) and related techniques. These approaches probe the local frequency-dependent dielectric response of a sample with spatial resolutions ranging from 10 to 100 nm, determined by the tip apex size. Tip-enhanced THz fields are also crucial for terahertz scanning tunneling microscopy (THz-STM), which detects tunneling currents induced by intense THz near fields, achieving an unprecedented spatial resolution of less than 0.1 nm. Unlike conventional THz near-field microscopy, THz-STM exploits the THz probe pulse-induced voltage rather than its frequency content to achieve spectroscopic contrast, making it applicable beyond low-energy excitations. The bandwidth of THz pulses is primarily used to endow atomic-resolution scanning probe microscopy with ultrafast temporal resolution, enabling the observation of local dynamics in materials and devices at the ångström scale with unprecedented precision. The summary of imaging techniques in THz-TDS is shown in Table 4.

**Table 4**

The summary of imaging techniques in THz-TDS.

| Techniques | Price | Signal type | Advantages | Disadvantages |
|---|---|---|---|---|
| THz-TDS | 6 (Cheapest) | Time-domain | High temporal resolution and providing amplitude and phase information | Single pixel imaging and complex systems |
| NLC | 5 | Time-domain | Fast imaging speed | Requiring high power laser and low SNR |
| TVSM | 7 | Time-domain | Fast imaging speed and stable system | Requiring spatial modulator and multiple encoded time intervals |
| THz-FPA | 3 | Frequency-domain | Fast imaging speed | No phase information |
| ACD, LTEM | 4 | Time-domain | High spatial resolution | Complex technology |
| THz-STM | 1 (Most expensive) | Time-domain | Atom-level resolution and accurate temporal resolution | Complex technology, only suitable for low energy excitation |
| s-SNOM | 2 | Frequency-domain | Nanometer-level resolution | Dependent on the size and performance of tip materials |

## 4. Artificial intelligence for IRT and THz techniques

Deep learning focuses on training artificial neural networks with multiple layers to automatically extract features and learn patterns from vast amounts of data. Inspired by the structure and function of the human brain, deep learning models, particularly deep neural networks (DNNs), have revolutionized fields such as computer vision, natural language processing, and speech recognition. In this section, we will review the applications of deep learning on IRT, THz, and spectroscopic techniques.

### *4.1. Infrared thermography*

During the 1990s, the researchers started to use neural networks as a thermographic nondestructive evaluation (TNDE) defect detector and classifier for analysis detection issues. In 1993, Prabhu et al. [161] applied two neural networks to detect subsurface flaw and estimate the corrosion. In 1997, Tretout et al. [162] used a neural network to train and distinguish the difference between the thermal curves of defect and sound area. In 1998, Maldague et al. [163] used a neural network to estimate the defect depth, which revealed the quantitative analysis based on neural networks is limited by the insufficient sampling frequency and the thermal diffusivity of materials themselves. In 2000, Vallerand et al. [164] proposed an innovative statistical processing method, which involves two different architectures of Perceptron neural network and Kohonen neural network for defect detection and characterization. In 2007, Benitez et al. [165] proposed a feature extraction of data based on DAC with artificial neural networks to identify the defects in composite materials. This neural network can effectively enhance the defect contrast comparing with original images. In 2012, Liu et al. [166] proposed a single hidden layer network to estimate the defect depth in LIT.

With the rapid development of graphics processing unit (GPU), advanced deep neural networks (DNNs) are beginning to replace the conventional artificial neural networks (ANNs). Duan et al. [167] combined long short-term memory recurrent neural network (LSTM-RNN) with TSR method to automatically classify common defects occurring including debonding, adhesive pooling, and liquid ingress in honeycomb materials. Some researchers employed (modified) U-Net neural networks for semantic segmentation of thermograms [168-170]. For automatic defect detection, advanced neural networks like region-based convolutional neural networks (R-CNN) and you only look once (YOLO) series were employed in many literatures [171,172]. In the case of denoising, unsupervised learning has been used in IRT. For instance, autoencoder and physical-informed neural networks (PINNs) were employed to enhance the defect contrast and reduce the background noise [173,174]. Super-resolution is a relatively new field in both computer vision and infrared thermography. The advanced generative adversarial network (GAN) was introduced to improve the spatial resolution, especially the defect edge [175].

### *4.2. Terahertz time-domain spectroscopy*

The deep learning techniques have been widely used to enhance the performance of THz technology through providing support for data processing. In 2002, Ferguson et al. [176] employed a quadratic classifier with linear filter-based feature extraction to address the challenge of slow acquisition speed of THz systems. In 2006, Zhong et al. [177] used a camera-like approach extracts absorption peaks for classification using minimum distance and neural network methods. In 2007, Yin et al. [178] applied support vector machine (SVM) for two- and multiclass classification. In 2009, Li et al. [179] used two artificial neural networks (organized mapping and multilayer perceptron) to identify weakly absorptive chemicals. In 2010, Hua et al. [180] combined partial least squares (PLS) and least-squares support vector machine methods to detect the pesticide residues. In 2015, Ge et al. [181] compared the detection effect using PLS, principal component regression, SVM and PCA-SVM for aflatoxin B1 in acetonitrile solutions. Zhan et al. [182] proposed several chemometric models including PLS and back propagation-artificial neural network for quantitatively detecting n-heptane volume ratios in n-heptane and n-octane mixtures. In 2019, Long et al. [183] proposed an effective and robust method for THz image super-resolution based on a deep CNN, which performs better than other super-resolution methods on the synthetic THz images. In 2020, Ljubenović et al. [184] applied convolutional neural networks (CNNs) to reduce THz beam shape effects for the first time. Experimental results show superior performance of CNNs compared to traditional model-based deblurring methods. In 2021, Wang et al. [185] presented three deep neural networks (DNNs) including long-short term memory recurrent neural network (LSTM-RNN) model, one-dimension CNN model, and a bidirectional LSTM-RNN model for predicting defect depths in glass fiber reinforced polymer (GFRP). The results show that one-dimension CNN performs better than the other two neural networks. In 2022, Hung et al. [186] proposed a THz deep learning computed tomography framework, which is a supervised CNN without predefined features and prior information. The extracted data-driven features are interpretable and be highly correlated to physic implication based on saliency maps. In 2024, Jiang et al. [187] combined THz-TDS with a faster R-CNN with coordinate attention (Fast R-CNN-CA), which achieved automatic inspection of the cultural heritage. Sun et al. [188] presented a physics-constrained transfer learning framework to improve the accuracy of terahertz thickness measurements of thermal barrier coatings. Zhu et al. [189] combined PCA with super-resolution generative adversarial networks to breakthrough the Abbe diffraction limit.

## 5. Conclusion

In conclusion, non-invasive inspection techniques operating in the mid-infrared to terahertz regions play a significant role in revealing surface and subsurface defects, detecting hidden drawings or inscriptions, and supporting material identification.

Infrared thermography has attracted increasing interest in the field of cultural heritage and has been successfully applied in practical inspection scenarios. However, one major limitation of this technique lies in its shallow detection depth, resulting from the gradient-driven nature of heat propagation and lateral thermal diffusion, which degrades the signal-to-noise ratio (SNR). While passive thermography can improve penetration depth by exploiting environmental thermal stimuli, it remains highly dependent on ambient conditions, often resulting in lower SNR compared to active methods.

Moreover, quantitative evaluation in passive thermography remains challenging due to the need for complex thermal modelling. Recent efforts have focused on the use of signal modulation strategies, such as chirped thermal excitations, to enhance detection depth. Nevertheless, further experimental validation and quantitative analysis are needed to assess the accuracy and practical feasibility of these approaches. Notably, very few studies have addressed the potential risks associated with thermal excitation.

While such effects may be negligible in industrial contexts, they cannot be disregarded when examining unique and irreplaceable cultural heritage. In addition, the infrared camera—an essential component of thermography systems—still faces technical limitations, including relatively low spatial resolution and sensitivity, along with high cost.

Terahertz imaging represents a relatively recent and rapidly evolving non-invasive diagnostic tool in both scientific and engineering domains. Positioned between photonics and electronics, the fundamental principles of terahertz photonics remain an active area of investigation. The development of effective imaging components—particularly sources and detectors—is being extensively pursued in the fields of physics and materials science. Despite its great potential, the widespread application of THz imaging still faces significant challenges, including the need for high-efficiency sources, highly sensitive detectors, cost-effective solutions, and simplified configurations suitable for large-scale experimental setups. Recent research efforts have primarily focused on mid-field imaging to enhance spatial resolution. However, current implementations remain distant from real-world or industrial applications. Although strategies such as deep learning and compressive sensing have been shown to offer some benefits, the overall improvement remains limited. As such, further progress in THz imaging is expected to depend largely on advancements in system design and integration.

This review began with a presentation of the fundamental principles of infrared thermography and terahertz imaging, highlighting their application in the non-destructive investigation of cultural heritage objects. Particular attention was given to real case studies involving authentic historical artifacts, in order to demonstrate the practical relevance of these techniques. Key contributions—including pioneering work, studies focused on physical mechanisms, and emerging approaches—have been emphasized.

Advanced signal processing techniques in both hardware and software have driven the development of non-invasive inspection and non-destructive testing. In infrared thermography, several classic but effective image processing algorithms were introduced and compared. For instance, PCA often achieves the best results in all excitation modalities but lacks physical information. PPT can provide the important depth information since the thermal diffusion length is related to the frequency. TSR can effectively reduce the temporal / frequency noise and compress the data storage. In THz-TDS, we discussed the signal processing algorithms for one-dimensional signals and analyzed the reasons leading to low-frequency blur and high-frequency noise.

Subsequently, we reviewed the application of artificial intelligence in these non-invasive inspection techniques from infrared region to terahertz region. For infrared thermography, all machine learning methods were introduced in order of time from 1990s. It is possible to find that all kinds of applications of deep learning in infrared thermography have been explored due to the fast development of artificial intelligence and graphics processing units (GPUs). For THz-TDS, since it has similar principles to hyperspectral imaging, the initial attempt is focused on classification of materials and chemical components. Then researchers combined advanced deep neural networks such as super-resolution networks and physics-constrained transfer learning techniques with THz-TDS techniques.

In summary, this work presents a critical and comprehensive review of the fundamental principles, experimental configurations, signal processing strategies, real-world applications, and potential developments in non-invasive imaging techniques spanning from the mid-infrared to the terahertz region. It is intended to serve as a reference for conservation scientists and engineers engaged in the application of advanced diagnostic methods for cultural heritage preservation.

This review is built upon the collective efforts of the research community in the fields of non-destructive testing and heritage science, with the aim of acknowledging and highlighting the significance of their contributions. By consolidating current knowledge and practical examples, it is hoped that this work will support further understanding and implementation of these techniques in the protection and study of cultural assets.

## Declaration of competing interest

The authors declare that they have known competing financial interests or personal relationships that could have appeared to influence the work reported in this paper.

## Acknowledgments

This work was supported by the National Key R&D Program of China (Grant n. 2023YEF0197800) and the Italian Ministry of University and Research (Grant n. PGR02110 – CUP: E13C24000350001), and Natural Sciences and Engineering Research Council (NSERC) of Canada through the Discovery and CREATE 'oN DuTy!' program (496439-2017), the Canada Research Chair in Multipolar Infrared Vision (MiViM).